\documentclass[12pt]{article}
\usepackage{amsmath}
\usepackage{graphicx}
\usepackage{color}
\usepackage{orcidlink}
\usepackage{caption}
\usepackage{subcaption}
\usepackage{setspace}
\usepackage{color}
\usepackage{hyperref}
\hypersetup{colorlinks=true, linkcolor=blue, citecolor=blue, urlcolor=blue}
\RequirePackage[numbers,sort&compress]{natbib}
\begin{document}
\baselineskip=22pt

\begin{center}
    {\large {\bf Stationary BTZ space-time in Ricci-Inverse and $f(\mathcal{R})$ gravity theories }}
\end{center}

\vspace{0.1cm}

\begin{center}
    {\bf Faizuddin Ahmed\orcidlink{0000-0003-2196-9622}}\footnote{\bf faizuddinahmed15@gmail.com}\\
    {\it Department of Physics, University of Science \& Technology Meghalaya, Ri-Bhoi, Meghalaya, 793101, India}\\
    \vspace{0.2cm}
    {\bf Abdelmalek Bouzenada\orcidlink{0000-0002-3363-980X}}\footnote{\textbf{abdelmalekbouzenada@gmail.com (Corresponding author)  }}\\
    {\it Laboratory of Theoretical and Applied Physics, Echahid Cheikh Larbi Tebessi University 12001, Algeria}
\end{center}

\vspace{0.2cm}

\begin{abstract}
In this paper, we explore a stationary BTZ space-time within the framework of modified gravity theory, specifically focusing on Ricci-inverse gravity. It is important to clarify that ``Ricci-inverse" refers to the inverse of the Ricci tensor, not the Ricci scalar. We consider a general class of this gravity theory, where the function $f$ is defined by $f(\mathcal{R}, \mathcal{A}, A^{\mu\nu}\,A_{\mu\nu})$, with $\mathcal{R}$ and $\mathcal{A}$ representing the Ricci and anti-curvature scalars, respectively and $A^{\mu\nu}$ is the anti-curvature tensor. We demonstrate that stationary BTZ space-time is a valid solution in this gravity theory, wherein the cosmological constant undergoes modifications due to the coupling constants. Moreover, we study another modified gravity theory known as $f(\mathcal{R})$-gravity and analyze the stationary BTZ space-time. Subsequently, we fully integrate the geodesic equations for BTZ space-time constructed within the Ricci-Inverse gravity, expressing the solutions in terms of elementary functions and compared with the GR result. We classify different types of geodesics, including null and time-like geodesics, under three conditions: (i) nonzero mass and angular momentum, $M \neq 0, J\neq 0$, (ii) massless BTZ space-time, $M=0$ and $J=0$, and (iii) $M=-1, J=0$, that is $AdS_3$-type, and analyze the results in modified gravity theories and compare with the general relativity case.
\end{abstract}

\vspace{0.1cm}

{\bf Keywords}: Modified gravity theories; BTZ space-time; Geodesics Equations

\vspace{0.1cm}

{\bf PACS number(s)}: 04.50.Kd; 04.20.Jb; 98.80.Es 

\section{Introduction}\label{s1}

The general theory of relativity (GTR) \cite{t2} redefined gravity as the curvature of space-time rather than a conventional force, offering profound insights into the structure of the universe. One of its most captivating predictions is the existence of black holes-regions where space-time curvature becomes so extreme that nothing, not even light, can escape. Over the years, various black hole solutions have been identified in familiar (3+1) dimensions, such as the Schwarzschild black hole for non-rotating, uncharged cases, and the Kerr black hole for rotating ones. Each of these solutions has contributed significantly to our understanding of gravitational collapse and event horizons \cite{r1,r2,r3,r4}. In standard general relativity, the Einstein field equations are expressed as follows:
\begin{equation}
R^{\mu\nu} - \frac{1}{2}\,\mathcal{R}\,g^{\mu\nu} +\Lambda\,g^{\mu\nu}= \frac{8\,\pi\,G}{c^4}\,T^{\mu\nu},
\end{equation}
where $R^{\mu\nu}$ is the contravariant Ricci tensor, $\mathcal{R}=g_{\mu\nu}\,R^{\mu\nu}$ is the Ricci scalar, $g^{\mu\nu}$ is the contravariant metric tensor, $\Lambda$ is the cosmological constant, $G$ is the gravitational constant, $c$ is the speed of light, and $T^{\mu\nu}$ is the stress-energy tensor.

However, black hole solutions are not confined to 3+1 dimensions; they have also been studied in both lower and higher-dimensional spacetimes, offering diverse theoretical landscapes. Notably, in lower dimensions, particularly in (2+1) spacetime, M. Ba\~{n}ados, C. Teitelboim, and J. Zanelli made a groundbreaking discovery in 1992, demonstrating that the Einstein-Maxwell equations with a negative cosmological constant yield a black hole solution, now famously known as the BTZ black hole \cite{r5,r6}. This solution is significant because it highlights unique features absent in higher-dimensional black holes. For instance, while the BTZ black hole exhibits mass, angular momentum, and charge similar to its 3+1 dimensional counterparts, it lacks a curvature singularity at the origin \cite{r7}. Additionally, the BTZ black hole can possess event and Cauchy horizons, especially in rotating or charged configurations, providing a rich structure for studying the causal boundaries of spacetime. The rotating BTZ black hole is described by a specific metric given by \cite{r8}:
\begin{equation}\label{BTZ_metric}
\mathrm{d}s^2 = -f(r)\mathrm{d}t^2 + \frac{1}{f(r)}\mathrm{d}r^2 + r^2\left[\mathrm{d}\phi-\Omega(r)\mathrm{d}t\right]^2~,
\end{equation}
where $f(r)$ and $\Omega(r)$ describe the gravitational potential and angular velocity, respectively. Despite its simplicity compared to higher-dimensional black holes, the BTZ black hole has become a vital tool in the exploration of lower-dimensional quantum gravity and holography, due to its solvability and the absence of singularities, shedding light on the fundamental nature of spacetime itself \cite{r9}.

Modified gravity theories represent a class of theoretical frameworks designed to address the limitations of Einstein's general relativity \cite{q1,q2,q3,q4}, particularly in explaining phenomena such as dark energy, dark matter, and quantum gravitational effects \cite{q5}. While general relativity has been remarkably successful in describing gravitational interactions on solar system scales and predicting phenomena like black hole formation and gravitational waves \cite{q6,q7,q8}, it faces challenges when accounting for the accelerated expansion of the universe attributed to dark energy, as well as the unseen matter responsible for the majority of a galaxy's mass-dark matter \cite{q9,q10}. Additionally, the integration of gravity with quantum mechanics remains a significant hurdle \cite{q11}.

In modified gravity theories, several extensions of general relativity have been proposed to explain cosmic phenomena such as dark energy and dark matter without invoking new forms of matter. Among these, $f(\mathcal{R})$-gravity \cite{m1,m2,m3,kk2} extends the gravitational action to a function of the Ricci scalar $\mathcal{R}$, while $f(\mathcal{G})$-gravity \cite{s3,s4,n3,kk8,kk3,kk4}  incorporates the Gauss-Bonnet invariant $\mathcal{G}$, an important topological term that captures higher-order curvature effects. The $ f(\mathcal{R}, \mathcal{G}) $-gravity \cite{s5,s6,ctp,n1,kk1}  considers both the Ricci scalar and the Gauss-Bonnet term, allowing the theory to address both early-universe inflation and late-time cosmic acceleration. Another notable extension is $ f(\mathcal{R}, T)$-gravity \cite{s7,k1,k2,kk6,kk10}, where the action depends on the Ricci scalar $\mathcal{R}$ and the trace of the stress-energy tensor $T$,  introducing direct interactions between matter and geometry. Similarly, $f(\mathcal{G}, T)$-gravity broadens this framework by including both the Gauss-Bonnet invariant and the trace $T$, facilitating a wider range of cosmological behaviors.

On a different front, $f(\mathcal{T})$-gravity \cite{s9,s10,n6} involves a function of the torsion scalar $\mathcal{T}$, positioning itself within the realm of teleparallel gravity, which explains gravitation through torsion rather than curvature. Another intriguing model is  $f(\mathcal{Q})$-gravity \cite{s11,s12,n5,kk9,kk11,kk14,kk15}, where $\mathcal{Q}$ represents the nonmetricity scalar, measuring the failure of space-time to be metric-compatible. This framework explores how spacetime geometry can deviate from general relativity through nonmetricity, offering a novel perspective on gravitational effects. More complex models, such as $ f(\mathcal{R}, \mathcal{G}, T) $-gravity \cite{s13,s14,n8,SS1} and $ f(\mathcal{Q}, T) $-gravity \cite{s15,s16,n4} incorporate multiple scalars (e.g., Ricci scalar, Gauss-Bonnet invariant, trace of the energy-momentum tensor, or nonmetricity scalar) to provide a richer set of dynamics and potentially explain a broader array of cosmic phenomena. Additionally, $ f(\mathcal{R}, \mathcal{L}_m) $-gravity \cite{s17,s18,n7}, where $\mathcal{L}_m$ is the matter Lagrangian. Some other studies in modified gravity theories were reported in Refs. \cite{kk5,kk7,kk13,PP1,PP2,PP3,PP4,PP5,PP6}.

Among many modified theories, Ricci-inverse gravity has attracted much attention in current times  \cite{ref1,ref2,ref3}. This approach modifies the Einstein field equations by introducing terms that are inversely proportional to the Ricci tensor, a quantity that encapsulates the curvature of space-time. The motivation behind Ricci-inverse modifications is to explore new dynamical possibilities for spacetime that may provide solutions to problems at both cosmological and quantum scales \cite{q17}. By altering gravitational dynamics in this manner, Ricci-inverse gravity has the potential to influence the large-scale structure of the universe, offering new explanations for the formation of cosmic structures and the behavior of dark energy \cite{q18,q19}. Furthermore, these modifications could yield novel black hole solutions that differ from those predicted by general relativity, thereby enhancing our understanding of high-energy gravitational phenomena and the conditions of the early universe \cite{q20}.

Ricci-Inverse gravity is a novel fourth-order gravity model \cite{ref1} which is founded on the introduction of a geometric entity called the anti-curvature scalar, $\mathcal{A}$. To be more specific, this anti-curvature scalar $\mathcal{A}$ is defined as the trace of the anti-curvature tensor $A^{\mu\nu}$ which is itself is the inverse of the Ricci tensor, $R^{-1}_{\mu\nu}$ or written as $A^{\mu\nu}\,R_{\mu\sigma}=\delta^{\nu}_{\sigma}$. In Ref. \cite{ref2}, authors considered three classes of Ricci-inverse gravity known as Class {\bf I} to Class {\bf III}-models. The Class {\bf I} model is defined by $f(\mathcal{R}, \mathcal{A})$, where $f$ is a function of the Ricci scalar $\mathcal{R}$, and the anti-curvature scalar $\mathcal{A}$, respectively. The Class-{\bf II} model is defined by $f(\mathcal{R}, A^{\mu\nu}\,A_{\mu\nu})$, where $f$ is a function of the Ricci scalar and the quadratic invariant of the anti-curvature tensor. The Class-{\bf III} model or general Class is defined by $f(\mathcal{R}, \mathcal{A}, A^{\mu\nu}\,A_{\mu\nu})$, where $f$ is a function of the Ricci scalar, the anti-curvature scalar, and the quadratic invariant of the anti-curvature tensor. Noted that Class-{\bf I} to Class-{\bf II} are the special cases of Class-{\bf III} model. In Ref. \cite{ref3}, authors have discussed in details the essential of modified gravity theories and compared with the general relativity theory (see, also Refs. \cite{ref6,ref7} for discussion of alternative theories). Recently, this Ricci-inverse theory has gained significant interest among researchers in current times (see, Refs. \cite{TQD,TQD2,EPJC,q14,q13,AM1,AM2,AM3,AM4,AM5,AM6,AM8,PLB,NA,CJPHY,EPJC3,EPJC4,EPJP,NPB,AOP,JCAP})

The significance of current study lies in its examination of stationary BTZ space-time within the framework of modified gravity theories, such as Ricci-Inverse gravity and $f(\mathcal{R})$-gravity. These approach enables us to investigate how modifications to the gravitational action, defined by the function $\mathcal{R} \to f(\mathcal{R}, \mathcal{A}, A^{\mu\nu}A_{\mu\nu})=(\mathcal{R} + \alpha_1\, \mathcal{R}^2+\alpha_2\, \mathcal{R}^3+\beta_1\, \mathcal{A}+\beta_2\, \mathcal{A}^2+\gamma\,A^{\mu\nu}\,A_{\mu\nu})$ for the Ricci-Inverse gravity, and $\mathcal{R} \to f(\mathcal{R})=(\mathcal{R} + \alpha_1\, \mathcal{R}^2+\alpha_2\, \mathcal{R}^3)$ for $f(\mathcal{R})$-gravity influence the stationary BTZ space-time properties. Subsequently, we study the geodesic equations on BTZ space-time within the framework of Ricci-Inverse gravity, expressing the solutions in terms of elementary functions and coupling constants. We classify different types of geodesics, null and time-like geodesics, under three conditions: (i) $M \neq 0, J\neq 0$, (ii) $M=-1$ and $J=0$, and (iii) $M=0=J$. We demonstrate that geodesic motions of both massive and massless test particles are influenced by the aforementioned modified gravity theories, and modifies the results compared to general relativity case.

The structure of this paper is organized as follows: In Section \ref{sec:2}, we explore stationary BTZ spacetime within the framework of Ricci-Inverse gravity theory, defined by the function $f\left(\mathcal{R}, \mathcal{A}, A^{\mu\nu}A_{\mu\nu}\right)$. We derive the solutions and compare them with those from general relativity, highlighting special cases of these solutions. In Section \ref{sec:3}, we study the same space-time model within the framework of another gravity theory known as $f(\mathcal{R})$ gravity and analyze the results. In Section \ref{sec:4}, we solve the geodesic equations, obtaining complete integrals for both null and time-like geodesics, and again compare these results with the general relativity cases. Finally, in Section \ref{sec:5}, we present our discussion and conclusions. The system of units are chosen, where $c=1=G$.

\section{Stationary BTZ space-time in Ricci-Inverse gravity}\label{sec:2}

We begin this section by introducing the three-dimensional BTZ black hole rotating metric by the following line-element in the "Schwarzschild" coordinates ($t, r, \psi$) \cite{r5,r6}
\begin{eqnarray}
ds^2=g_{tt}\,dt^2+2\,g_{t\psi}\,dt\,d\psi+g_{rr}\,dr^2+g_{\psi\psi}\,d\psi^2,\label{1}
\end{eqnarray}
where different coordinates are in the ranges $-\infty < t=x^0 < +\infty$, $0 < r=x^1 < +\infty$, and $\psi=x^3 \sim \psi+2\,n\,\pi$ ($n=\pm\,1,\pm\,2,....$). The metric tensor $g_{\mu\nu}$ and its contravariant form $g^{\mu\nu}$ are given by $(\mu,\nu=0,1,2)$
\begin{eqnarray}
g_{\mu\nu}&=&\left(\begin{array}{ccc}
    M+\Lambda\,r^2 & 0 & -\frac{J}{2}\\
     0 & \frac{1}{\Big(-M-\Lambda\,r^2+\frac{J^2}{4\,r^2}\Big)} & 0\\
     -\frac{J}{2} & 0 & r^2
\end{array}\right),\nonumber\\
g^{\mu\nu}&=&\left(\begin{array}{ccc}
     \frac{1}{\Big(M+\Lambda\,r^2-\frac{J^2}{4\,r^2}\Big)} & 0 & \frac{J}{2\,r^2\Big(M+\Lambda\,r^2-\frac{J^2}{4\,r^2}\Big)} \\
     0 & \Big(-M-\Lambda\,r^2+\frac{J^2}{4\,r^2}\Big) & 0\\
     \frac{J}{2\,r^2\Big(M+\Lambda\,r^2-\frac{J^2}{4\,r^2}\Big)} & 0 & \frac{M+\Lambda\,r^2}{r^2\,\Big(M+\Lambda\,r^2-\frac{J^2}{4\,r^2}\Big)}
\end{array}\right).\label{3}
\end{eqnarray}

The two constants of integration, $M$ and $J$, correspond to the conserved charges associated with asymptotic invariance under time displacements (representing mass) and rotational invariance (representing angular momentum), respectively. Here, $\Lambda=-\ell^{-2}$ represents the cosmological constant, with $\ell$ being the radius of curvature. It is noteworthy that the radius of curvature, denoted as $\ell=(-\Lambda)^{-1/2}$, establishes the length scale required to induce a horizon in a theory where the mass is dimensionless.

The covariant Ricci tensor $R_{\mu\nu}$ and its contravariant form $R^{\mu\nu}$ for the metric (\ref{1}) will be given by
\begin{eqnarray}
R_{\mu\nu}&=&\left(\begin{array}{ccc}
     2\,\Lambda\,(M+\Lambda\,r^2) & 0 & -J\,\Lambda\\
     0 & \frac{2\,\Lambda}{\Big(-M-\Lambda\,r^2+\frac{J^2}{4\,r^2}\Big)} & 0\\
     -J\,\Lambda & 0 & 2\,\Lambda\,r^2
\end{array}\right),\nonumber\\
R^{\mu\nu}&=&\left(\begin{array}{ccc}
     \frac{2\,\Lambda}{\Big(M+\Lambda\,r^2-\frac{J^2}{4\,r^2}\Big)} & 0 & \frac{J\,\Lambda/r^2}{\Big(M+\Lambda\,r^2-\frac{J^2}{4\,r^2}\Big)}\\
     0 & 2\,\Lambda\,\Big(-M-\Lambda\,r^2+\frac{J^2}{4\,r^2}\Big) & 0\\
     \frac{J\,\Lambda/r^2}{\Big(M+\Lambda\,r^2-\frac{J^2}{4\,r^2}\Big)} & 0 & \frac{2\,\Lambda}{r^2}\,\frac{M+\Lambda\,r^2}{\Big(M+\Lambda\,r^2-\frac{J^2}{4\,r^2}\Big)}
\end{array}\right).\label{4}
\end{eqnarray}
Finally, the Ricci scalar is given by
\begin{equation}
    \mathcal{R}=g_{\mu\nu}\,R^{\mu\nu}=6\,\Lambda\,.\label{5}
\end{equation}

One can easily show that this three-dimensional space-time (\ref{1}) satisfies the vacuum Einstein field equations with a negative cosmological constant given by
\begin{equation}
    G_{\mu\nu}+\Lambda\,g_{\mu\nu}=0,\quad \Lambda<0\,.\label{6}
\end{equation}

Now, we focus on Ricci-Inverse gravity theory by considering metric (\ref{1}) as background model. It is important to note that in Ricci-Inverse gravity theory, the determinant of the Ricci tensor $R_{\mu\nu}$ associated with any metric must be non-zero. Moreover, we would like to mention here that the Ricci-Inverse gravity doesn't mean the inverse of the Ricci scalar $\mathcal{R}$. In our current model, we observe that the determinant of the Ricci tensor $R_{\mu\nu}$ for the line element (\ref{1}) is non-zero, as given by:
\begin{eqnarray}
    \mbox{det}\,(R_{\mu\nu})=-8\,r^2\,\Lambda^3\,.\label{7}
\end{eqnarray}
Since the determinant of the Ricci tensor $R_{\mu\nu}$ is nonzero which indicates the existence of a symmetric anti-curvature tensor $A^{\mu\nu}$ defined by
\begin{eqnarray}
    A^{\mu\nu}=R^{-1}_{\mu\nu}=\frac{\mbox{adj}\,(R_{\mu\nu})}{\mbox{det}\,(R_{\mu\nu})}.\label{8}
\end{eqnarray}

For space-time (\ref{1}), we obtain this anti-curvature tensor $A^{\mu\nu}$ and its covariant form given by
\begin{eqnarray}
A^{\mu\nu}&=&\left(\begin{array}{ccc}
     \frac{1}{2\,\Lambda}\,\frac{1}{\Big(M+\Lambda\,r^2-\frac{J^2}{4\,r^2}\Big)} & 0 & \frac{J}{4\,r^2\,\Lambda}\,\frac{1}{\Big(M+\Lambda\,r^2-\frac{J^2}{4\,r^2}\Big)}\\
     0 &  \frac{1}{2\,\Lambda}\,\Big(-M-\Lambda\,r^2+\frac{J^2}{4\,r^2}\Big) & 0\\
     \frac{J}{4\,r^2\,\Lambda}\,\frac{1}{\Big(M+\Lambda\,r^2-\frac{J^2}{4\,r^2}\Big)} & 0 & \frac{1}{2\,\Lambda\,r^2}\,\frac{M+\Lambda\,r^2}{\Big(M+\Lambda\,r^2-\frac{J^2}{4\,r^2}\Big)}
\end{array}\right),\nonumber\\
A_{\mu\nu}&=&\left(\begin{array}{ccc}
     \frac{M+\Lambda\,r^2}{2\,\Lambda} & 0 &-\frac{J}{4\,\Lambda}\\
     0 & \frac{1}{2\,\Lambda}\,\frac{1}{\Big(-M-\Lambda\,r^2+\frac{J^2}{4\,r^2}\Big)} & 0\\
     -\frac{J}{4\,\Lambda} & 0 & \frac{r^2}{2\,\Lambda}
\end{array}\right).\label{9}
\end{eqnarray}
The anti-curvature scalar is given by
\begin{equation}
    \mathcal{A}=g_{\mu\nu}\,A^{\mu\nu}=\frac{3}{2\,\Lambda}\,.\label{10}
\end{equation}
The quadratic invariant of the anti-curvature tensor is given by
\begin{equation}
    A^{\mu\nu}\,A_{\mu\nu}=\frac{3}{4\,\Lambda^2}.\label{11}
\end{equation}

We now introduce anti-curvature tensor $A^{\mu\nu}$ into the Einstein-Hilbert action of the system. Therefore, the action that describes the Ricci-inverse gravity in general class is given by \cite{ref1,ref2}
\begin{eqnarray}
    \mathcal{S}= \int \sqrt{-g}\left[f(\mathcal{R}, \mathcal{A}, A^{\mu\nu}\,A_{\mu\nu})-2\,\Lambda\right]\,dt\,d^2x++{\cal S}_m,\label{13}
\end{eqnarray}

where ${\cal S}_m$ is the action of matter Lagrangian, and other symbols have their usual meanings. This action (\ref{13}) can be rewritten as follows:
\begin{eqnarray}
    \mathcal{S}= \int \sqrt{-g}\left[f(g_{\mu\nu}\,R^{\mu\nu}, g_{\mu\nu}\,A^{\mu\nu},A^{\mu\nu}\,A_{\mu\nu}) -2\,\Lambda_m\right]\,dt\,d^2x++{\cal S}_m.\label{14}
\end{eqnarray}

By varying the action (\ref{14}) with respect to the metric tensor $g_{\mu\nu}$, the modified field equations is given by  
\begin{equation}\label{15}
    -\frac{1}{2}\, f\, g^{\mu \nu} + f_{\mathcal{R}} \, R^{\mu \nu}-f_{\mathcal{A}} \, A^{\mu \nu} - 2\,f_{A^2}\, A^{\rho \nu }A^{\mu}_{\rho} +P^{\mu \nu}+ M^{\mu \nu} + U^{\mu \nu} + \Lambda_m g^{\mu \nu} =T^{\mu \nu},
\end{equation}
where $\Lambda_m$ is the cosmological constant in this  Ricci-inverse gravity theory, $T^{\mu \nu}$ is the energy-momentum tensor, and other tensors are as follows:
\begin{eqnarray}
    P^{\mu \nu} &=& g^{\mu \nu} \nabla ^2 \, f_{\mathcal{R}}-\nabla ^{\mu} \nabla^{\nu}\, f_{\mathcal{R}} \, ,\label{16}\\ 
    M^{\mu \nu} &=& g^{\rho \mu}\nabla _{\alpha } \nabla _{\rho } (f_{\mathcal{A}}\, A_{\sigma}^{ \alpha}\, A^{\nu \sigma}) - \frac{1}{2}\, \nabla ^2 (f_{\mathcal{A}}\,A^{\mu}_{\sigma}\, A^{\nu \sigma}) - \frac{1}{2}\, g^{\mu \nu}\, \nabla _{\alpha} \nabla _{ \rho} ( f_{\mathcal{A}}\,A_{\sigma}^{ \alpha}\, A^{\rho \sigma})\, ,\label{17}\\
 \nonumber  U^{\mu \nu} &=& g^{\rho \nu}\,\nabla _{\alpha} \nabla _{\rho}(f_{A^2}\,A_{\sigma \kappa}\,A^{\sigma \alpha}\,A^{\mu \kappa})-\nabla ^{2}(f_{A^2}\,A_{\sigma \kappa}\,A^{\sigma \mu}\,A^{\nu \kappa})
  -g^{\mu \nu}\,\nabla _{\alpha} \nabla _{\rho}(f_{A^2}\,A_{\sigma \kappa}\,A^{\sigma \alpha}\,A^{\rho \kappa})\nonumber\\
  &+& 2\,g^{\rho \nu}\,\nabla _{\rho} \nabla _{\alpha}(f_{A^2}\,A_{\sigma \kappa}\,A^{\sigma \mu}\,A^{\alpha \kappa})
   - g^{\rho \nu}\,\nabla _{\alpha} \nabla _{\rho}(f_{A^2}\,A_{\sigma \kappa}\,A^{\sigma \mu}\,A^{\alpha \kappa})\, .\label{18} 
\end{eqnarray}
Here various symbols are defined by
\begin{equation}
    f_{\mathcal{R}}=\partial f/\partial \mathcal{R},\quad\quad f_{\mathcal{A}}=\partial f/\partial \mathcal{A},\quad\quad f_{A^2}=\partial f/\partial (A^{\mu\nu}\,A_{\mu\nu})\,.\label{19}
\end{equation}

Let us consider the function $f(\mathcal{R},\mathcal{A},A^{\mu \nu}\,A_{\mu \nu})$ in general Class of Ricci-Inverse gravity theory to be the following form:
\begin{equation} \label{20}
    f(\mathcal{R},\mathcal{A},A^{\mu \nu}\,A_{\mu \nu}) =\mathcal{R}+{\alpha_1\,\mathcal{R}^2}+\alpha_2\,\mathcal{R}^3+\beta_1\, \mathcal{A}+\beta_2\,\mathcal{A}^2+\gamma\,A^{\mu \nu}\,A_{\mu \nu}, 
\end{equation}
where $\alpha_i, \beta_i$\quad ($i=1,2$) and $\gamma$ being arbitrary constants. Using this function, one can find the following symbols:
\begin{eqnarray}\label{21}
    f_{\mathcal{R}}=1+2\,\alpha_1\,\mathcal{R}+3\,\alpha_2\,\mathcal{R}^2,\quad\quad f_{\mathcal{A}}=\beta_1+2\,\beta_2\,\mathcal{A},\quad\quad
    f_{A^2}=\gamma.
\end{eqnarray}

We known that both the Ricci scalar $\mathcal{R}$ given in Eq. (\ref{6}) and the anti-curvature scalar $\mathcal{A}$ provided in Eq. (\ref{10}) are constant. Hence, the expression of $f_{\mathcal{R}}$ given in (\ref{21}) is also a constant. Consequently, we find that the tensor $P^{\mu\nu}$ in Eq. (\ref{16}) must vanish.

Therefore, the modified field equations (\ref{15}) using (\ref{21}) reduces to the following form:
\begin{eqnarray}\label{22}
    &&-\frac{1}{2}\,\left(\mathcal{R}+\alpha_1\,\mathcal{R}^2+\alpha_2\,\mathcal{R}^3+\beta_1 \,\mathcal{A}+\beta_2\,\mathcal{A}^2+\gamma\,A^{\mu \nu}\,A_{\mu \nu}\right)\,g^{\mu \nu}+ (1+2\,\alpha_1\,\mathcal{R}\nonumber\\
    &&+3\,\alpha_2\,\mathcal{R}^2)\, R^{\mu \nu}
    -(\beta_1+\beta_2\,\mathcal{A})\,A^{\mu \nu}-2\,\gamma\, A^{\rho \nu }\,A^{\mu}_{\rho}+M^{\mu \nu} +U^{\mu \nu} + \Lambda g^{\mu \nu} =T^{\mu \nu},\quad
\end{eqnarray}
where various tensor quantities now reduces to as follows:  
\begin{eqnarray}
    P^{\mu \nu}&=&0,\nonumber\\
    M^{\mu \nu}&=&(\beta_1+\beta_2\,\mathcal{A})\,\Big[g^{\rho \mu}\nabla _{\alpha } \nabla _{\rho } (A_{\sigma}^{ \alpha}\, A^{\nu \sigma}) - \frac{1}{2}\, \nabla ^2 (A^{\mu}_{\sigma}\, A^{\nu \sigma}) - \frac{1}{2}\, g^{\mu \nu}\, \nabla _{\alpha} \nabla _{ \rho} (A_{\sigma}^{ \alpha}\, A^{\rho \sigma})\Big],\nonumber\\
    U^{\mu \nu}&=&\gamma\,\Big[g^{\rho \nu}\,\nabla _{\alpha} \nabla _{\rho}(A_{\sigma \kappa}\,A^{\sigma \alpha}\,A^{\mu \kappa})-\nabla ^{2}(A_{\sigma \kappa}\,A^{\sigma \mu}\,A^{\nu \kappa})
  -g^{\mu \nu}\,\nabla _{\alpha} \nabla _{\rho}(A_{\sigma \kappa}\,A^{\sigma \alpha}\,A^{\rho \kappa})\nonumber\\
  &+& 2\,g^{\rho \nu}\,\nabla _{\rho} \nabla _{\alpha}(A_{\sigma \kappa}\,A^{\sigma \mu}\,A^{\alpha \kappa})
   - g^{\rho \nu}\,\nabla _{\alpha} \nabla _{\rho}(A_{\sigma \kappa}\,A^{\sigma \mu}\,A^{\alpha \kappa})\Big].\label{23} 
\end{eqnarray}

By substituting the anti-curvature tensor provided in Eq. (\ref{9}), the anti-curvature scalar from Eq. (\ref{10}), and the metric tensor from Eq. (\ref{3}) into the modified field equations outlined in Eq. (\ref{22}), we derive the following non-zero components of the energy-momentum tensor: 

\begin{small}
\begin{eqnarray}
\nonumber &&T^{tt}=\frac{1}{\left(-M-\Lambda\,r^2+\frac{J^2}{4\,r^2}\right)}\,\frac{\left[15\,\beta_2+7\,\gamma+10\,\beta_1\,\Lambda+8\,\Lambda^3- 
 48\,\alpha_1\,\Lambda^4-864\,\alpha_2\,\Lambda^5- 
 8\,\Lambda^2\,\Lambda_m\right]}{8\,\Lambda^2},\\
 \nonumber &&T^{t\psi}=\frac{J}{2\,r^2}\,\frac{1}{\left(-M-\Lambda\,r^2+\frac{J^2}{4\,r^2}\right)}\,\frac{\left[15\,\beta_2+7\,\gamma+10\,\beta_1\,\Lambda+8\,\Lambda^3- 
 48\,\alpha_1\,\Lambda^4-864\,\alpha_2\,\Lambda^5- 
 8\,\Lambda^2\,\Lambda_m\right]}{8\,\Lambda^2},\\
    \nonumber &&T^{rr}=-\left(-M-\Lambda\,r^2+\frac{J^2}{4\,r^2}\right)\,\frac{\left[15\,\beta_2+7\,\gamma+10\,\beta_1\,\Lambda+8\,\Lambda^3- 
 48\,\alpha_1\,\Lambda^4-864\,\alpha_2\,\Lambda^5- 
 8\,\Lambda^2\,\Lambda_m\right]}{8\,\Lambda^2},\\
    &&T^{\psi\psi}=\frac{M/r^2+\Lambda}{\left(-M-\Lambda\,r^2+\frac{J^2}{4\,r^2}\right)}\,\frac{\left[15\,\beta_2+7\,\gamma+10\,\beta_1\,\Lambda+8\,\Lambda^3- 
 48\,\alpha_1\,\Lambda^4-864\,\alpha_2\,\Lambda^5- 
 8\,\Lambda^2\,\Lambda_m\right]}{8\,\Lambda^2}.\label{24}
\end{eqnarray}\end{small}

By solving the modified field equations given in Eq. (\ref{24}) for a vacuum, where the energy-momentum tensor is $T^{\mu \nu}=0$, and after simplifying the resulting expressions, we obtain the effective cosmological constant in this modified gravity theory given by: 
\begin{equation}
    \Lambda_m=\Lambda-6\,\alpha_1\,\Lambda^2-108\,\alpha_2\,\Lambda^3+\frac{5\,\beta_1}{4\,\Lambda}+\frac{15\,\beta_2}{8\,\Lambda^2}+\frac{7\,\gamma}{8\,\Lambda^2}.\label{25}
\end{equation}

\begin{figure}[ht!]
    \centering
    \includegraphics[width=0.71\linewidth]{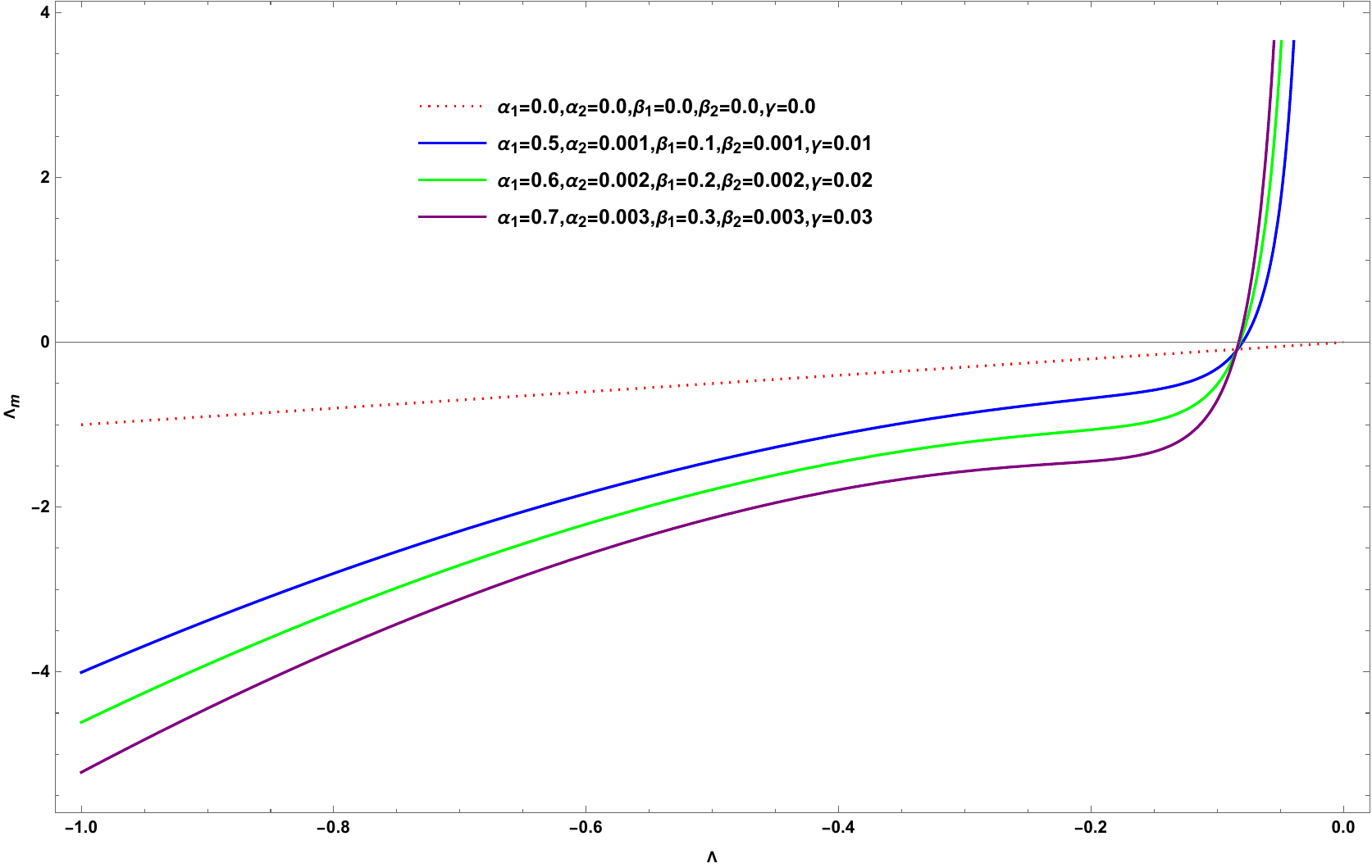}
    \caption{The effective cosmological constant $\Lambda_m$ as a function of $\Lambda$ for different values of the coupling constants.}
    \label{fig:1}
\end{figure}

We observe that the effective cosmological constant is modified by the coupling constants $(\alpha_i,\beta_i,\gamma)$ in comparison to the result in General Relativity (GR). Figure \ref{fig:1} illustrates this modified cosmological constant as a function of the cosmological constant in GR. The dotted red line represents the GR result, while the solid lines correspond to different values of the coupling constants in modified gravity.

Below, we discuss some special case of this modified gravity theory.

\vspace{0.5cm}
\begin{center}
{\bf Case A:\quad Class-{\bf I} model of RI-gravity or $f(\mathcal{R},\mathcal{A})$-gravity}\label{s2-2}
\end{center}

In this case, we discuss modified gravity corresponds to $\alpha_1=0=\alpha_2,\quad\gamma=0$. Therefore, the function (\ref{20}) reduces to the following form:
\begin{equation} \label{bb1}
    f=\mathcal{R}+\beta_1\,\mathcal{A}+\beta_2\,\mathcal{A}^2  
\end{equation}
which contains both the Ricci and anti-curvature scalars, respectively. Therefore, we have Class-{\bf I} models of Ricci-Inverse gravity theory or $f(\mathcal{R},\mathcal{A})$-gravity theory.

Following the previous procedure, after deriving the modified field equations and simplification for zero energy-momentum tensor results the following effective cosmological constant given by
\begin{equation}
    \Lambda_m=\Lambda+\frac{5\,\beta_1}{4\,\Lambda}+\frac{15\,\beta_2}{8\,\Lambda^2}\label{bb2}
\end{equation}
which reduces to those result obtained in \cite{CJPHY} for $\beta_2=0$.

From the above analysis, we conclude that for zero energy-momentum tensor, the anti-curvature scalar $\mathcal{A}^i$, where $i=1,2,.....$ introduced into the Einstein-Hilbert action influences the usual cosmological constant $\Lambda$ and gets modification by the coupling constants $\beta_i$.

The effective cosmological constant $\Lambda_m$ remains negative provided we have the following constraint
\begin{equation}
    \beta_2 < -\frac{8\,\Lambda^2}{15}\left(\Lambda+\frac{5\,\beta_1}{4\,\Lambda}\right).\label{bb3}
\end{equation}

\vspace{0.5cm}

\begin{center}
{\bf Case B:\quad Class-{\bf II} model of RI-gravity or $f(\mathcal{R}, A^{\mu\nu}\,A_{\mu\nu})$-gravity}\label{s2-3}
\end{center}

In this case, we discuss modified gravity corresponds to $\alpha_1=0=\alpha_2,\quad \beta_1=0=\beta_2$. Therefore, the function (\ref{20}) reduces to the following form:
\begin{equation} \label{cc1}
    f=\mathcal{R}+\gamma\,A^{\mu\nu}\,A_{\mu\nu}  
\end{equation}
which contains the Ricci scalar and quadratic invariant of the anti-curvature tensor. Therefore, we have Class-{\bf II} model of Ricci-Inverse gravity or $f(\mathcal{R},A^{\mu\nu}\,A_{\mu\nu})$-gravity theory.

Following the previous procedure, after deriving the modified field equations and simplification for zero energy-momentum tensor results the following effective cosmological constant given by
\begin{equation}
    \Lambda_m=\Lambda+\frac{7\,\gamma}{8\,\Lambda^2}<0,\label{cc2}
\end{equation}
provided we have the following constraint
\begin{equation}
    0 < \gamma<-\frac{8\,\Lambda^3}{7}.\label{ccc2}
\end{equation}

From this analysis, we conclude that for zero energy-momentum tensor, quadratic invariant of anti-curvature tensor $(A^{\mu\nu}\,A_{\mu\nu})^i$, where $i=1,2,.....$ introduced into the Einstein-Hilbert action influences the usual cosmological constant $\Lambda$ and gets modification by the coupling constants $\gamma_i$.

The stationary BTZ metric therefore in Ricci-Inverse gravity in general Class $f(\mathcal{R},\mathcal{A},A^{\mu\nu}\,A_{\mu\nu})$ is described by the following line-element
\begin{eqnarray}
ds^2=g_{tt}\,dt^2+2\,g_{t\psi}\,dt\,d\psi+g_{rr}\,dr^2+g_{\psi\psi}\,d\psi^2,\label{special-metric}
\end{eqnarray}
where the metric tensor $g_{\mu\nu}$ now is given by 
\begin{eqnarray}
g_{\mu\nu}=\left(\begin{array}{ccc}
     -(-M-\Lambda_m\,r^2) & 0 & -\frac{J}{2}  \\
     0 & \frac{1}{\Big(-M-\Lambda_m\,r^2+\frac{J^2}{4\,r^2}\Big)} & 0\\
     -\frac{J}{2} & 0 & r^2
\end{array}\right),\label{special-metric2}
\end{eqnarray}
where the effective cosmological constant $\Lambda_m$ is now expressed in Eq. (\ref{25}). Depending on the values of $M$ and $J$ various space-time emerge from the BTZ metric (\ref{special-metric})--(\ref{special-metric2}) which are summarized in Table \ref{tab:1}.

\begin{table}[ht!]
    \centering
    \begin{tabular}{c|c}
    \hline
     M-J regions    &  Space-times\\ [2.5ex] 
     \hline\hline
     $M>0$ and $|J|>\frac{M}{\sqrt{-\Lambda_m}}$     & Black Holes (BH) \\ [2.5ex] 
     \hline
     $M>0$ and $|J|=\frac{M}{\sqrt{-\Lambda_m}}$     & Extreme BH \\ [2.5ex] 
     \hline
     $M>0$ and $|J|<-\frac{M}{\sqrt{-\Lambda_m}}$     & Naked Singularities (NS) \\ [2.5ex] 
     \hline
     $M>0$ and $|J|=-\frac{M}{\sqrt{-\Lambda_m}}$     & Extreme NS \\ [2.5ex] 
     \hline
     $M=0$ and $J=0$     & Massless BTZ BH \\ [2.5ex]
     \hline
     $M=-1$ and $J=0$     & AdS$_3$ vacuum \\ [2.5ex] 
     \hline
    \end{tabular}
    \caption{BTZ space-times for different $M$ and $J$.}
    \label{tab:1}
\end{table}

\section{Stationary BTZ space-time in $f(\mathcal{R})$-gravity }\label{sec:3}

In this section, we study $f(\mathcal{R})$-gravity theory by considering stationary BTZ space-time (\ref{1}) as background model. In this theory, the Ricci scalar $\mathcal{R}$ is replaced by a function of itself, $\mathcal{R} \to f(\mathcal{R})$, in the Einstein-Hilbert action. Therefore, the Lagrangian that describes this $f(\mathcal{R})$-gravity theory is given by
\begin{eqnarray}
    S= \int dt\,d^2x\, \sqrt{-g}\left[f(\mathcal{R})-2\,\Lambda_m+{\cal L}_m\right].\label{ss1}
\end{eqnarray}

Varying the action with respect to the metric tensor results the modified field equations for $f(\mathcal{R})$-gravity with the cosmological constant given by 
\begin{equation}\label{ss2}
    f_{\mathcal{R}}\,R^{\mu \nu}-\frac{1}{2}\, f(\mathcal{R})\, g^{\mu \nu}-\nabla^{\mu}\, \nabla^{\nu} \, f_{\mathcal{R}} + g^{\mu \nu}\, \nabla^{\alpha}\, \nabla_{\alpha} \, f_{\mathcal{R}} + \Lambda _m\, g^{\mu \nu}=T^{\mu \nu}.
\end{equation}

In this analysis, we choose the function $f(\mathcal{R})$ to be the following form 
\begin{equation} \label{ss4}
    f(\mathcal{R})=\mathcal{R}+{\alpha_1\,\mathcal{R}^2}+\alpha_2\,\mathcal{R}^3 
\end{equation}
which reduces to GR for zero coupling constants $\alpha_1=0=\alpha_2$. The presence of higher order curvature terms thus modifies the dynamics of the system. 

By substituting this function $f$ given in (\ref{ss4}), the metric tensor specified in (\ref{3}), and the Ricci scalar provided in (\ref{5}) into the modified field equations in (\ref{ss2}), we obtain the following non-zero components:
\begin{eqnarray}
    \nonumber &&T^{tt}=\frac{1}{\left(-M-\Lambda\,r^2+\frac{J^2}{4\,r^2}\right)}\,(\Lambda- 
 6\,\alpha_1\,\Lambda^2-108\,\alpha_2\,\Lambda^3-\Lambda_m),\\
 \nonumber &&T^{t\psi}=\frac{J}{2\,r^2}\,\frac{1}{\left(-M-\Lambda\,r^2+\frac{J^2}{4\,r^2}\right)}\,(\Lambda- 
 6\,\alpha_1\,\Lambda^2-108\,\alpha_2\,\Lambda^3-\Lambda_m),\\
    \nonumber &&T^{rr}=-\left(-M-\Lambda\,r^2+\frac{J^2}{4\,r^2}\right)\,(\Lambda- 
 6\,\alpha_1\,\Lambda^2-108\,\alpha_2\,\Lambda^3-\Lambda_m),\\
    &&T^{\psi\psi}=\frac{M/r^2+\Lambda}{\left(-M-\Lambda\,r^2+\frac{J^2}{4\,r^2}\right)}\,(\Lambda- 
 6\,\alpha_1\,\Lambda^2-108\,\alpha_2\,\Lambda^3-\Lambda_m).\label{ss5}
\end{eqnarray}

By solving the field equations (\ref{ss5}) for zero energy-momentum tensor, $T^{\mu\nu}=0$, we find, after simplification, that the effective cosmological constant is given as follows:
\begin{equation}
    \Lambda_m=\Lambda-6\,\alpha_1\,\Lambda^2-108\,\alpha_2\,\Lambda^3.\label{ss6}
\end{equation}
that reduces to GR result for $\alpha_1=0=\alpha_2$.

The effective cosmological constant thus remain negative $\Lambda_m<0$ provided we have the following constraint
\begin{equation}
    \alpha_2<\frac{1}{108\,\Lambda^2}-\frac{\alpha_1}{18\,\Lambda}.\label{ss7}
\end{equation}

From expression (\ref{ss6}), we see that stationary BTZ space-time is a valid vacuum solution in $f(R)$-gravity theory too in addition to Ricci-Inverse gravity, where the cosmological constant gets modification by the coupling constants.

We have generated Figure \ref{fig:2} showing the influences of the coupling constant on the effective cosmological constant in this $f(\mathcal{R})$-gravity.

\begin{center}
\begin{figure}[ht!]
\begin{centering}
\subfloat[$\alpha_2=0.001$]{\centering{}\includegraphics[scale=0.62]{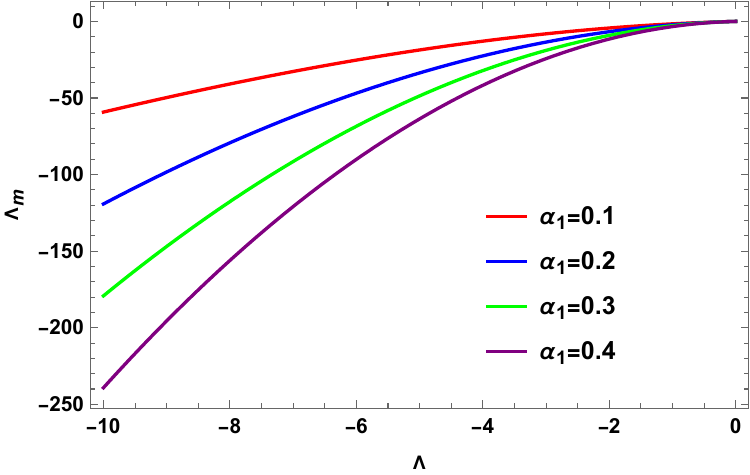}}\quad\quad
\subfloat[$\alpha_1=0.1$]{\centering{}\includegraphics[scale=0.62]{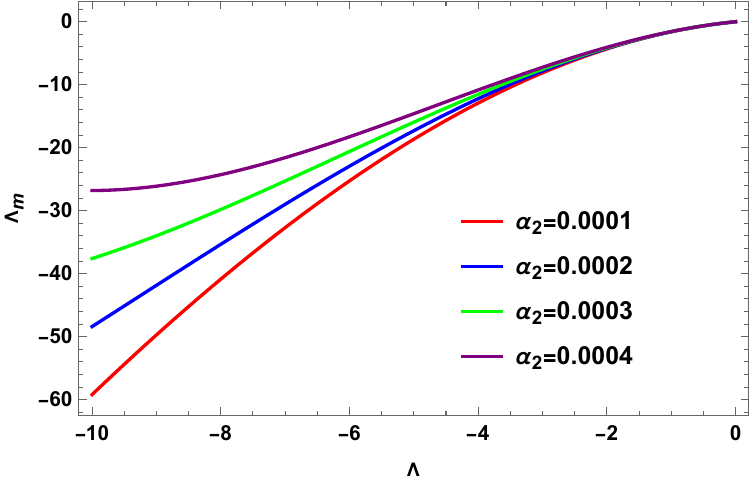}}
\hfill\\
\begin{centering}
\subfloat[]{\centering{}\includegraphics[scale=0.65]{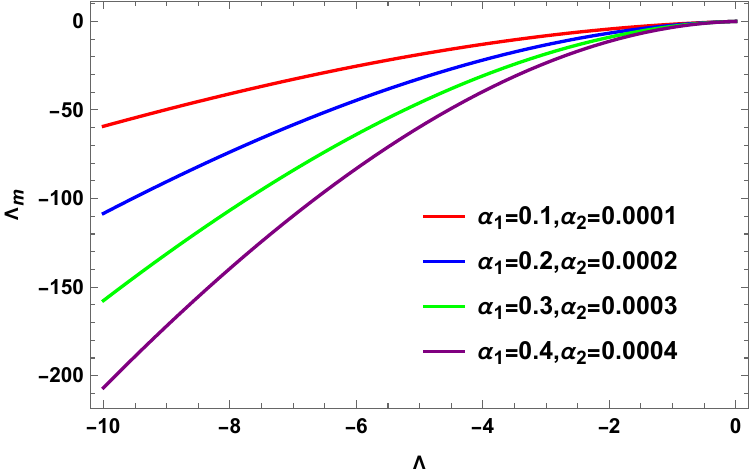}}
\end{centering}
\centering{}\caption{The effective cosmological constant $\Lambda_m$ of Eq. (\ref{ss6}) within $f(\mathcal{R})$-gravity.}\label{fig:2}
\end{centering}
\end{figure}
\par\end{center}

\section{Geodesics Equations: Comparison of modified gravity theories  and GR}\label{sec:4}

In this section, we study the geodesic motions of test particles around this BTZ space-time given by the metric (\ref{special-metric}) and compare the results with those obtained in general relativity.

The geodesics equations are given by
\begin{equation}
    \ddot{x}^{\lambda}+\Gamma^{\lambda}_{\mu\nu}\,\dot{x}^{\mu}\,\dot{x}^{\nu}=0,\label{A1}
\end{equation}
where dot represents derivative w. r. t. affine parameter $\tau$.

The metric tensor $g_{\mu\nu}$ for the space-time (\ref{special-metric}) depends on the coordinate $r$ only, and is independent of the coordinate $t, \psi$. Therefore, there exists two Killing vector $\partial_{t}=\frac{\partial}{\partial t}$ and $\partial_{\psi}=\frac{\partial}{\partial \psi}$. The corresponding constants of motion with respect to the parameter $\tau$ can be derived using the relation $k=g_{\mu\nu}\,\xi^{\mu}\,\frac{dx^{\nu}}{d\tau}$. These constants $(\mathrm{E},\mathrm{L})$ are given by
\begin{eqnarray}
    -\mathrm{E}=g_{tt}\,\dot{t}+g_{t\psi}\,\dot{\psi},\nonumber\\
    \mathrm{L}=g_{t\psi}\,\dot{t}+g_{\psi\psi}\,\dot{\psi}.\label{A2}
\end{eqnarray}
After simplification, we get
\begin{eqnarray}
    &&\dot{t}=\frac{g_{t\psi}\,\mathrm{L}+g_{\psi\psi}\,\mathrm{E}}{g^2_{t\psi}-g_{tt}\,g_{\psi\psi}}=\frac{\mathrm{E}\,r^2-J\,\mathrm{L}/2}{r^2\,\left(-M-\Lambda_m\,r^2+\frac{J^2}{4\,r^2}\right)},\nonumber\\
    &&\dot{\psi}=-\frac{g_{tt}\,\mathrm{L}+g_{t\psi}\,\mathrm{E}}{g^2_{t\psi}-g_{tt}\,g_{\psi\psi}}=\frac{(-M-\Lambda_m\,r^2)\,\mathrm{L}+J\,\mathrm{E}/2}{r^2\,\left(-M-\Lambda_m\,r^2+\frac{J^2}{4\,r^2}\right)}.\label{A3}
\end{eqnarray}
It is worth mentioning that the denominators of $\dot{t}$ and $\dot{\psi}$ vanish at the BH horizons $r=r_h$, while for Naked singularities (NS) they are positive definite. Consequently, the geodesics around a NS are drastically different from those in the BH case.

The Lagrangian of the system is defined by
\begin{equation}
    \mathcal{L}=\frac{1}{2}\,g_{\mu\nu}\,\frac{dx^{\mu}}{d\tau}\,\frac{dx^{\nu}}{d\tau}.\label{A5}
\end{equation}

Using the metric tensor (\ref{special-metric}), we find
\begin{eqnarray}
    g_{tt}\,\dot{t}^2+2\,g_{t\psi}\,\dot{t}\,\dot{\psi}+g_{rr}\,\dot{r}^2+g_{\psi\psi}\,\dot{\psi}^2=\epsilon,\label{A6}
\end{eqnarray}
where $\epsilon=0$ for null geodesics and $-1$ for time-like geodesics.

Substituting metric tensor components and after simplification, we can write
\begin{eqnarray}
    \dot{r}^2=\epsilon\,\left(-M-\Lambda_m\,r^2+\frac{J^2}{4\,r^2}\right)+\mathrm{E}^2+\Lambda_m\,\mathrm{L}^2+\left(M\,\mathrm{L}^2-J\,\mathrm{E}\,\mathrm{L}\right)/r^2.\label{A7}
\end{eqnarray}
It is worth mentioning that for black holes, we have $M>0$ and naked singularities $M<0$.

The effective potential of the system is given by
\begin{eqnarray}
    V_{eff} (r)&=&-\epsilon\,\left(-M-\Lambda_m\,r^2+\frac{J^2}{4\,r^2}\right)-\Lambda_m\,\mathrm{L}^2-\left(M\,\mathrm{L}^2-J\,\mathrm{E}\,\mathrm{L}\right)/r^2, \label{A8}
\end{eqnarray}
where $\Lambda_m$ is given in Eq. (\ref{25}) in Ricci-Inverse gravity. We see that the effective potential of the system is influenced by the coupling constants $(\alpha_i,\beta_i,\gamma)$ involved in the modified cosmological constant (\ref{25}) in this gravity theory. These coupling constants alter the effective potential of the system compared to the result in General Relativity.

We have generated Figure \ref{fig:3}, which illustrates the behavior of the effective potential for null geodesics in both Ricci-Inverse gravity and General Relativity (GR). In panel (a) of Figure \ref{fig:3}, the dotted red line representing $\mathrm{L}=1$ corresponds to the GR case, where the parameters are set to $\alpha_i=0$, $\beta_i=0$, and $\gamma=0$. In contrast, the purple line for $\mathrm{L}=1$ represents the Ricci-inverse gravity scenario, with the coupling constants set as $\alpha_1=0.1$, $\alpha_2=0.001$, $\beta_1=0.1$, $\beta_2=0.001$, and $\gamma=0.01$. The other lines in this panel depict different values of  $\mathrm{L}$ while maintaining the same coupling constants. In panel (b) of Figure \ref{fig:3}, the lines correspond to various values of the coupling constants while keeping $\mathrm{L}=1$ fixed. This setup allows for a comparative analysis of the effective potential for null geodesics in both GR and Ricci-Inverse gravity theories. 

We have generated Figure \ref{fig:4}, which illustrates the behavior of the effective potential for time-like geodesics in both Ricci-Inverse gravity and General Relativity (GR). The explanation of this is analogue to the previous Figure \ref{fig:3}.

\begin{center}
\begin{figure}[ht!]
\begin{centering}
\subfloat[$\Lambda=-0.1$]{\centering{}\includegraphics[scale=0.62]{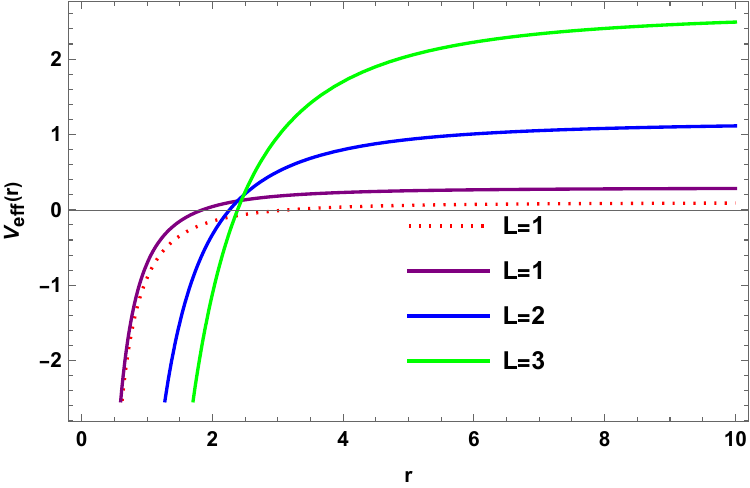}}\quad\quad
\subfloat[$\mathrm{L}=1$, $\Lambda=-0.1$]{\centering{}\includegraphics[scale=0.62]{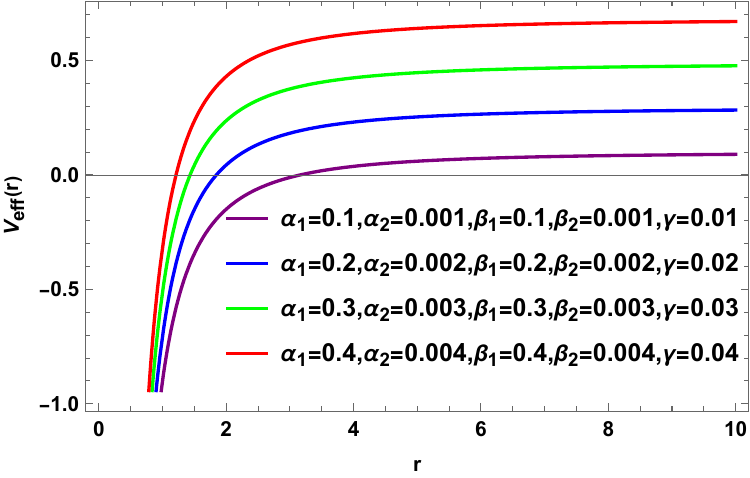}}
\centering{}\caption{The effective potential for null geodesics in BTZ space-time within the framework of Ricci-Inverse gravity. Here, we set $M= 2$, $J = 1$, $\mathrm{E}=1$.}\label{fig:3}
\hfill\\
\subfloat[]{\centering{}\includegraphics[scale=0.62]{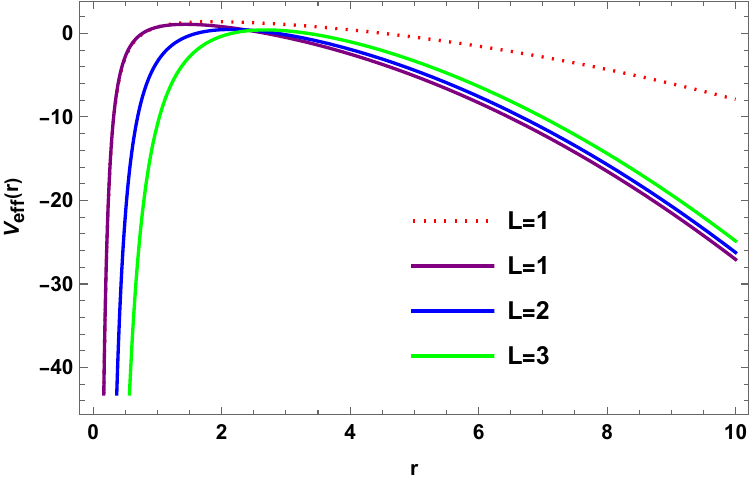}}\quad\quad
\subfloat[$\mathrm{L}=1$]{\centering{}\includegraphics[scale=0.62]{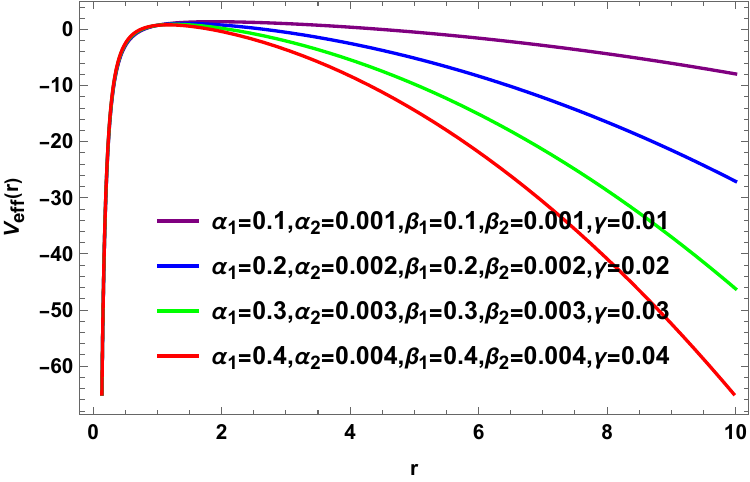}}\quad
\centering{}\caption{The effective potential for time-like geodesics in BTZ space-time within the framework of Ricci-Inverse gravity. Here, we set $M= 2$, $J = 1$, $\mathrm{E}=1$, $\Lambda=-0.1$.}\label{fig:4}
\end{centering}
\end{figure}
\par\end{center}

We have generated Figure \ref{fig:5}, which illustrates the behavior of the effective potential for null geodesics in both $f(\mathcal{R})$ gravity and General Relativity (GR). In panel (a) of Figure \ref{fig:5}, the dotted red line representing $\mathrm{L}=1$ corresponds to the GR case, where the parameters are set to $\alpha_1=0=\alpha_2$. In contrast, the purple line for $\mathrm{L}=1$ represents $f(\mathcal{R})$ gravity scenario, with the coupling constants set as $\alpha_1=0.8$, $\alpha_2=0.001$. The other lines in this panel depict different values of $\mathrm{L}$ while maintaining the same coupling constants.

In panel (b) of Figure \ref{fig:5}, the lines correspond to various values of the coupling constants while keeping $\mathrm{L}=1$ fixed. This setup allows for a comparative analysis of the effective potential for null geodesics in both GR and $f(\mathcal{R})$ gravity theories. 

We have generated Figure \ref{fig:6}, which illustrates the behavior of the effective potential for time-like geodesics in $f(\mathcal{R})$ gravity and General Relativity (GR). The explanation of this is analogue to the previous Figure \ref{fig:5}. 

\begin{center}
\begin{figure}[ht!]
\begin{centering}
\subfloat[$\Lambda=-0.1$]{\centering{}\includegraphics[scale=0.62]{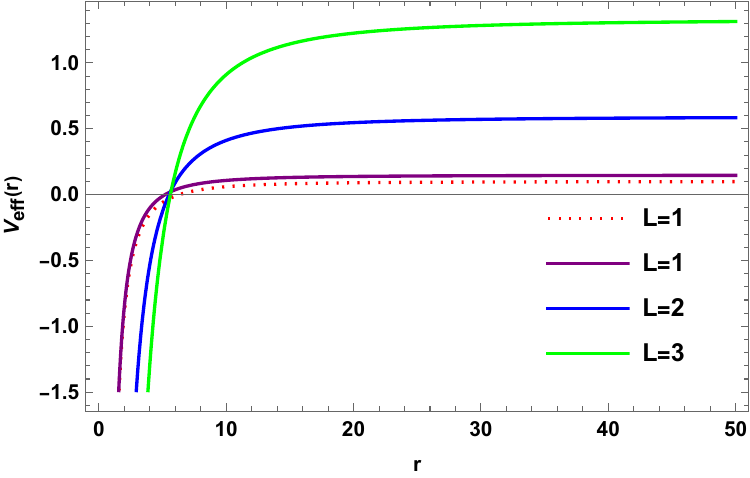}}\quad\quad
\subfloat[$\mathrm{L}=1$, $\Lambda=-0.1$]{\centering{}\includegraphics[scale=0.62]{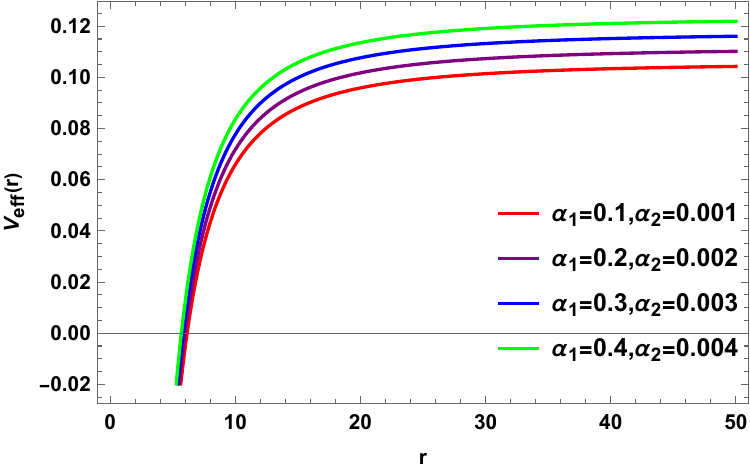}}
\centering{}\caption{The effective potential for null geodesics on BTZ space-time within the framework of $f(\mathcal{R})$ gravity. Here, we set $M=5$, $J=1$, $\mathrm{E}=1$.}\label{fig:5}
\hfill\\
\subfloat[]{\centering{}\includegraphics[scale=0.62]{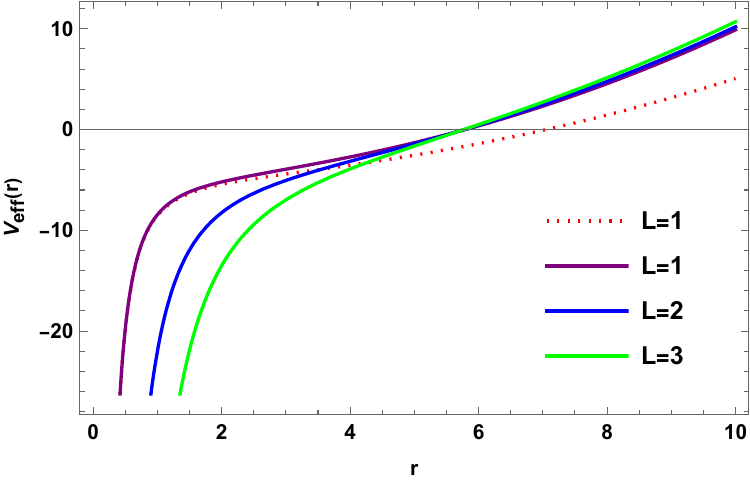}}\quad\quad
\subfloat[$\mathrm{L}=1$, $\Lambda=-0.1$]{\centering{}\includegraphics[scale=0.62]{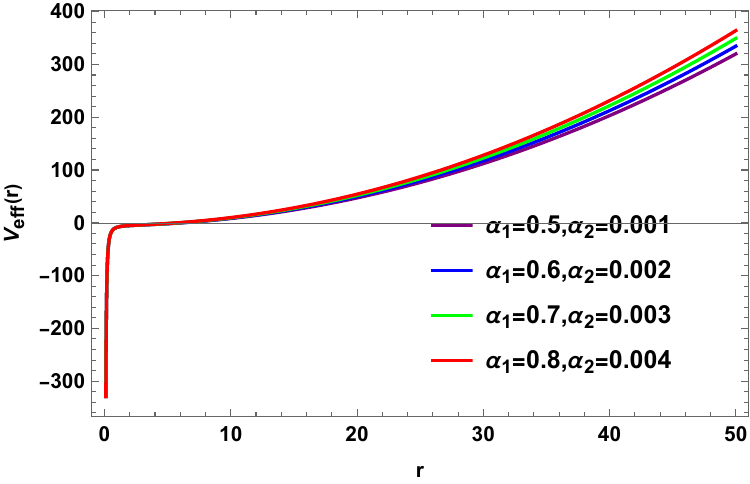}}\quad
\centering{}\caption{The effective potential for time-like geodesics on BTZ space-time within the framework of $f(\mathcal{R})$ gravity. Here, we set $M=5$, $J=1$, $\mathrm{E}=1$.}\label{fig:6}
\end{centering}
\end{figure}
\par\end{center}

Defining the following
\begin{equation}
    r^2\,\left(-M-\Lambda_m\,r^2+\frac{J^2}{4\,r^2}\right)=(-\Lambda_{m})\,\left(-r^2\,\Lambda_m+\mathrm{b}^2_{+}\right)\,\left(-r^2\,\Lambda_m+\mathrm{b}^2_{-}\right),\label{A9}
\end{equation}
where
\begin{equation}
    \mathrm{b}_{\pm}=\frac{1}{2}\,\left(\sqrt{-M+J\,\sqrt{-\Lambda_m}} \pm \sqrt{-M-J\,\sqrt{-\Lambda_m}}\right).\label{A10}
\end{equation}
We can rewrite the radial equation (\ref{A7})
\begin{eqnarray}
    (r\,\dot{r})^2&=&-\epsilon\,\Lambda_{m}\,\left(-r^2\,\Lambda_m+\mathrm{b}^2_{+}\right)\,\left(-r^2\,\Lambda_m+\mathrm{b}^2_{-}\right)+r^2\,(\mathrm{E}^2+\Lambda_m\,\mathrm{L}^2)-(\mathrm{b}^2_{+}+\mathrm{b}^2_{-})\,\mathrm{L}^2\nonumber\\
    &-&\frac{2\,\mathrm{b}_{+}\,\mathrm{b}_{-}\,\mathrm{E}\,\mathrm{L}}{\sqrt{-\Lambda_m}}.\label{A11}
\end{eqnarray}

\vspace{0.5cm}
{\bf Case A: Null Geodesics}
\vspace{0.5cm}

In this part, we study the null geodesics of the above system. Defining impact parameter
\begin{equation}
    \beta=\frac{\mathrm{L}}{\mathrm{E}}.\label{A12}
\end{equation}
The radial equation (\ref{A11}) for null geodesics $\epsilon=0$ will reduce to
\begin{eqnarray}
    r^2\,\dot{r}^2=\Bigg[r^2\,(1+\beta^2\,\Lambda_m)-(\mathrm{b}^2_{+}+\mathrm{b}^2_{-})\,\beta^2-\frac{2\,\beta\,\mathrm{b}_{+}\,\mathrm{b}_{-}}{\sqrt{-\Lambda_m}}\Bigg]\,\frac{\mathrm{L}^2}{\beta^2}.\label{A13}
\end{eqnarray}
Defining a new variable $u=\frac{1}{2}\,r^2$, we can write the above equation
\begin{equation}
    \dot{u}=\sqrt{\mathcal{A}\,u-\mathcal{B}},\label{A14}
\end{equation}
where
\begin{equation}
    \mathcal{A}=2\,(1+\beta^2\,\Lambda_m)\,\frac{\mathrm{L}^2}{\beta^2},\quad\quad \mathcal{B}=\Bigg[(\mathrm{b}^2_{+}+\mathrm{b}^2_{-})\,\beta^2+\frac{2\,\beta\,\mathrm{b}_{+}\,\mathrm{b}_{-}}{\sqrt{-\Lambda_m}}\Bigg]\,\frac{\mathrm{L}^2}{\beta^2}.\label{A15}
\end{equation}
The solution of Equation (\ref{A14}) is given by
\begin{equation}
    r(\tau)=\sqrt{\left(\frac{1}{\beta^2}+\Lambda_m\right)\,\mathrm{L}^2\,\tau^2+\frac{\left(-M+\frac{J}{\beta}\right)}{\left(\frac{1}{\beta^2}+\Lambda_m\right)}}.\label{A16}
\end{equation}

\vspace{0.5cm}
{\bf Case B: Time-like Geodesics}
\vspace{0.5cm}

In this part, we study the time-like geodesics of the system discussed earlier. The radial equation (\ref{A11}) for time-like  geodesics $\epsilon=-1$ and defining $u=\frac{1}{2}\,r^2$ will reduce to
\begin{eqnarray}
    \dot{u}^2=\Lambda^{2}_m\,\left(-4\,\Lambda_m\,u^2-2\,M\,u+\frac{J^2}{4}\right)+\mathcal{A}\,u-\mathcal{B}=f(u).\label{A17}
\end{eqnarray}
Time-like geodesics exist in the regions where the quadratic function of the RHS of Eq. (\ref{A17}) defined in the domain $u\geq 0$, is non-negative. In cases (a) $\mathcal{B}<0$ and (b) $\mathcal{A}>0$ and $\mathcal{B}=0$, the quadratic function $f(u)$ is non-negative in the interval $0 \leq u \leq u_{\pm}$, where
\begin{equation}
    u_{\pm}=\frac{1}{4\,(-\Lambda_m)}\,\left(-\frac{\mathcal{A}}{2\,\Lambda^2_{m}}+M\pm\sqrt{\Delta}\right),\label{A18}
\end{equation}
where we defined $\Delta$ as follows:
\begin{equation}
    \Delta=\left(M-\frac{\mathcal{A}}{2\,\Lambda^2_{m}}\right)^2+4\,\Lambda_m\,\left(\frac{J^2}{4}-\frac{\mathcal{B}}{\Lambda^2_{m}}\right).\label{A19}
\end{equation}

\subsection{Geodesics Equations on Massless BTZ Space-time}\label{subsec:1}

In this part, we will study the geodesics equation of the metric (\ref{special-metric}) under a very special case corresponds to zero mass, $M=0$, and the angular momentum, $J=0$, respectively. This scenario is called massless BTZ geometry and, therefore, the metric (\ref{special-metric}) can be written as
\begin{eqnarray}
ds^2=-(-\Lambda_m)\,r^2\,dt^2+\frac{dr^2}{(-\Lambda_m)\,r^2}+r^2\,d\psi^2.\label{special-metric3}
\end{eqnarray}

The geodesics equations for the space-time (\ref{special-metric3}) are given by
\begin{eqnarray}
    &&\dot{t}=\frac{\mathrm{E}}{(-\Lambda_m)\,r^2},\nonumber\\
    &&\dot{\psi}=\frac{\mathrm{L}}{r^2},\nonumber\\
    &&\dot{r}^2+(-\epsilon\,r^2+\mathrm{L}^2)\,(-\Lambda_m)=\mathrm{E}^2.\label{A20}
\end{eqnarray}
In that scenario, the effective potential of the system is given by
\begin{equation}
    V_{eff} (r)=\left(-\Lambda+6\,\alpha_1\,\Lambda^2+108\,\alpha_2\,\Lambda^3-\frac{5\,\beta_1}{4\,\Lambda}-\frac{15\,\beta_2}{8\,\Lambda^2}-\frac{7\,\gamma}{8\,\Lambda^2}  \right)\,\left(-\epsilon\,r^2+\mathrm{L}^2\right).\label{A21}
\end{equation}

\begin{center}
\begin{figure}[ht!]
\begin{centering}
\subfloat[$\Lambda=-0.1$]{\centering{}\includegraphics[scale=0.62]{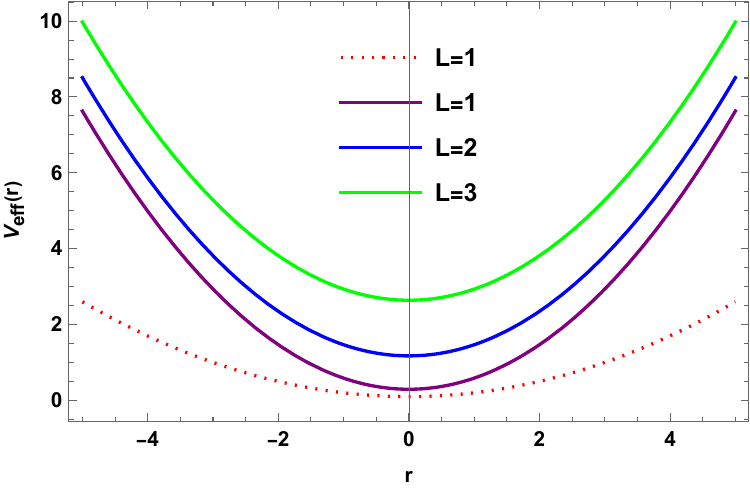}}\quad\quad
\subfloat[$\Lambda=-0.1$,$\mathrm{L}=1$]{\centering{}\includegraphics[scale=0.62]{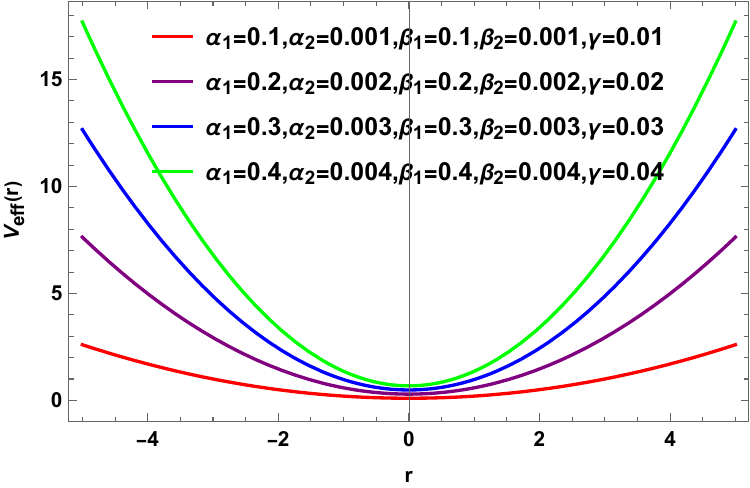}}\quad
\centering{}\caption{The effective potential for time-like geodesics on massless BTZ space-time within the framework of Ricci-Inverse gravity.}\label{fig:7}
\hfill\\
\subfloat[$\Lambda=-0.1$]{\centering{}\includegraphics[scale=0.62]{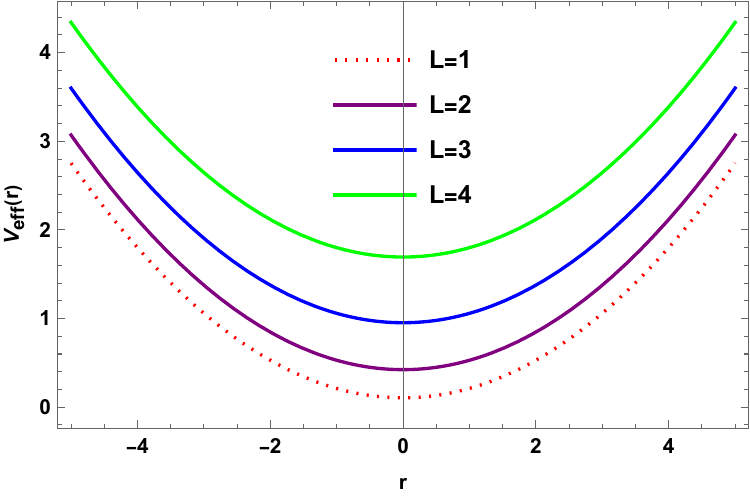}}\quad\quad
\subfloat[$\Lambda=-0.1$,$\mathrm{L}=1$]{\centering{}\includegraphics[scale=0.62]{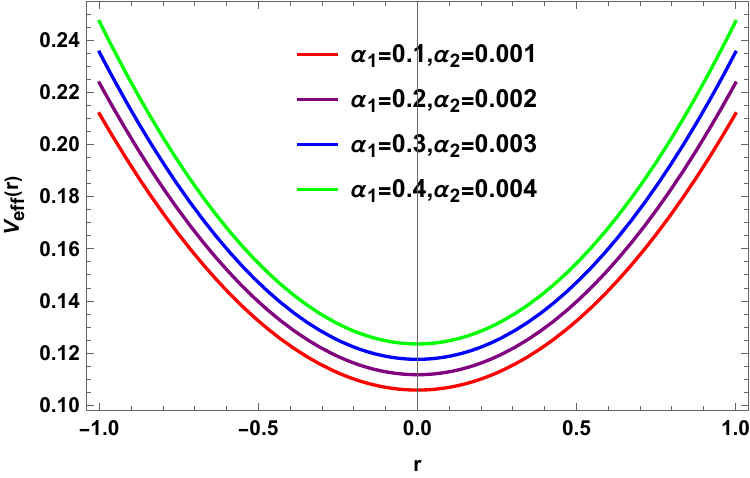}}\quad
\centering{}\caption{The effective potential for time-like geodesics on massless BTZ space-time within the framework of $f(\mathcal{R})$ gravity.}\label{fig:8}
\end{centering}
\end{figure}
\par\end{center}

We have generated Figure \ref{fig:7}, which illustrates the behavior of the effective potential for time-like geodesics in massless BTZ space-time within the framework of Ricci-Inverse gravity and General Relativity (GR). In panel (a) of Figure \ref{fig:7}, the dotted red line representing $\mathrm{L}=1$ corresponds to the GR case, where the parameters are set to $\alpha_i=0$, $\beta_i=0$, and $\gamma=0$. In contrast, the purple line for $\mathrm{L}=1$ represents the Ricci-Inverse gravity scenario, with the coupling constants set as $\alpha_1=0.1$, $\alpha_2=0.001$, $\beta_1=0.1$, $\beta_2=0.001$, and $\gamma=0.01$. The other lines in this panel depict different values of 
$\mathrm{L}$ while maintaining the same coupling constants. In panel (b) of Figure \ref{fig:8}, the lines correspond to various values of the coupling constants while keeping $\mathrm{L}=1$ fixed. This setup allows for a comparative analysis of the effective potential for null geodesics in both GR and Ricci-Inverse gravity theories by considering massless BTZ space-time. 

Similarly, we have depicted in Figure \ref{fig:8} the effective potential for time-like geodesics in massless BTZ space-time within the framework of $f(\mathcal{R})$ gravity for various values of $\mathrm{L}$ and the coupling constant $\alpha_1,\alpha_2$.

In the case of null geodesics $\epsilon=0$, we obtain
\begin{equation}
    r(\tau)=\pm\,\sqrt{\frac{\mathcal{A}}{2}}\,\tau+r_0,\quad\quad \mathcal{A}=2\,\left(\frac{1}{\beta^2}+\Lambda_m\right)\,\mathrm{L}^2, \label{A22}
\end{equation}
where $r_0>0$ is an arbitrary integration constant and $\mathcal{A}>0$. The range of the affine parameter is $0 \leq \tau < \infty$ for the upper sign, and $-\infty < \tau \leq r_0/\sqrt{\frac{\mathcal{A}}{2}}$  for the lower sign. For the case $\frac{1}{\beta^2}=-\Lambda_m$, which implies $\mathcal{A}=0$, one can find $r(\tau)=r_0$, which is a constant (circular orbit). The time component can be obtained as
\begin{equation}
    \frac{\dot{t}}{\dot{\psi}}=\frac{1}{(-\Lambda_m)\,\beta}\Rightarrow t(\tau)=\frac{1}{(-\Lambda_m)\,\beta}\,\psi(\tau).\label{time}
\end{equation}
We summarized null geodesics in Table \ref{tab:2}.

\begin{table}[ht!]
    \centering
    \begin{tabular}{c|c|c}
    \hline
    Case &  Range of $r$ & $r(\tau)$, $\psi(\tau)$\\[1ex]
    \hline
    $\epsilon=0$, $\frac{1}{\beta^2}>-\Lambda_m$ & $0 \leq r < \infty$ & $\begin{array}{c}
         r(\tau)=\pm\,\mathrm{L}\,\sqrt{\frac{1}{\beta^2}+\Lambda_m}\,\tau+r_0
         \hfill\\  [2ex] 
        \psi(\tau)=\mp\,\frac{1}{\mathrm{L}\,\sqrt{\frac{1}{\beta^2}+\Lambda_m}\,\left(\pm\,\mathrm{L}\,\sqrt{\frac{1}{\beta^2}+\Lambda_m}\,\tau+r_0\right)} 
    \end{array}$\\[5ex]
    \hline
    $\epsilon=0$, $\frac{1}{\beta^2}=-\Lambda_m$ & $r$ \mbox{constant and arbitrary} & $\begin{array}{c}
         r(\tau)=r_0
         \hfill\\  [2ex]
         \psi(\tau)=\frac{\mathrm{L}}{r^2_{0}}\,\tau
    \end{array}$\\[5ex]
    \hline 
\end{tabular}
\caption{Null Geodesic equations for massless BTZ space-time}
\label{tab:2}
\end{table}

For the time-like geodesics $\epsilon=-1$, we find from Eq. (\ref{A20}) as follow:
\begin{equation}
    \frac{\dot{r}}{\sqrt{r^2+\kappa^2}}=\sqrt{-\Lambda_m}\,,\label{A23}
\end{equation}
where
\begin{equation}
    \kappa^2=\frac{\mathcal{A}}{2\,(-\Lambda_m)}=\frac{1}{(-\Lambda_m)}\,\left(\frac{1}{\beta^2}+\Lambda_m\right)\,\mathrm{L}^2.\label{A24}
\end{equation}
The radial solution is given by
\begin{eqnarray}
r(s)=\left\{\,
\begin{array}{cc}
     \kappa\,\sinh \left(\sqrt{-\Lambda_m}\,s\right), &\quad \mbox{if} \quad  \kappa^2>0\Rightarrow \frac{1}{\beta^2}>-\Lambda_m,\\
     r_0\,\exp\left(\pm\,\sqrt{-\Lambda_m}\,s\right), &\quad \mbox{if} \quad  \kappa^2=0\Rightarrow \frac{1}{\beta^2}=-\Lambda_m,\\ 
     \kappa\,\cosh \left(\sqrt{-\Lambda_m}\,s\right), &\quad \mbox{if} \quad  \kappa^2<0\Rightarrow \frac{1}{\beta^2}<-\Lambda_m.
\end{array}
\right.
\label{A25}
\end{eqnarray}

Hence, the angular component from Eq. (\ref{A20}) is given by
\begin{eqnarray}
\psi(s)=\left\{\,
\begin{array}{cc}
     -\frac{\mathrm{L}}{\sqrt{-\Lambda_m}\,\kappa^2}\,\coth \left(\sqrt{-\Lambda_m}\,s\right), &\quad \mbox{if} \quad  \kappa^2>0\Rightarrow\frac{1}{\beta^2}>-\Lambda_m,
     \hfill\\  [2ex]
     \mp\,\frac{\mathrm{L}}{2\,r^2_{0}\,\sqrt{-\Lambda_m}}\,\exp\left(\mp\,2\,\sqrt{-\Lambda_m}\,s\right), &\quad \mbox{if} \quad  \kappa^2=0\Rightarrow \frac{1}{\beta^2}=-\Lambda_m,
     \hfill\\  [2ex] 
     \frac{\mathrm{L}}{\sqrt{-\Lambda_m}\,\kappa^2}\,\tanh \left(\sqrt{-\Lambda_m}\,s\right), &\quad \mbox{if} \quad  \kappa^2<0\Rightarrow\frac{1}{\beta^2}<-\Lambda_m.
\end{array}
\right.
\label{A26}
\end{eqnarray}
And the time component is given by
\begin{equation}
    t(s)=\frac{1}{(-\Lambda_m)\,\beta}\,\psi(s).\label{A27}
\end{equation}
We summarized time-like geodesics in Table \ref{tab:3}.

\begin{table}[ht!]
    \centering
    \begin{tabular}{c|c|c}
    \hline
    Case &  Range of $r$ & $r(s)$, $\psi(s)$, $t(s)$ \\[1ex]
    \hline
    $\epsilon=-1$, $\frac{1}{\beta^2}>-\Lambda_m$ & $0 \leq r < \infty$ & $\begin{array}{c}
         r(s)=\kappa\,\sinh \left(\sqrt{-\Lambda_m}\,s\right)
         \hfill\\  [2ex]
        \psi(s)=-\frac{\mathrm{L}}{\sqrt{-\Lambda_m}\,\kappa^2}\,\coth \left(\sqrt{-\Lambda_m}\,s\right)
        \hfill\\  [2ex]
        t(s)=-\frac{\mathrm{L}}{(-\Lambda_m)^{3/2}\,\beta\,\kappa^2}\,\coth \left(\sqrt{-\Lambda_m}\,s\right)
    \end{array}$\\[6ex]
    \hline 
    $\epsilon=-1$, $\frac{1}{\beta^2}=-\Lambda_m$ & $0 < r < \infty$  & $\begin{array}{c}
         r(s)=r_0\,\exp\left(\pm\,\sqrt{-\Lambda_m}\,s\right)
         \hfill\\ [2ex]
         \psi(s)=\mp\,\frac{\mathrm{L}}{2\,r^2_{0}\,\sqrt{-\Lambda_m}}\,\exp\left(\mp\,2\,\sqrt{-\Lambda_m}\,s\right)
         \hfill\\ [2ex]
         t(s)=\mp\,\frac{\mathrm{L}}{2\,r^2_{0}\,\beta\,(-\Lambda_m)^{3/2}}\,\exp\left(\mp\,2\,\sqrt{-\Lambda_m}\,s\right)
    \end{array}$\\[6ex]
    \hline 
    $\epsilon=-1$, $\frac{1}{\beta^2}<-\Lambda_m$ & $\kappa \leq r < \infty$ & $\begin{array}{c}
         r(s)=\kappa\,\cosh \left(\sqrt{-\Lambda_m}\,s\right)
         \hfill\\  [2ex]
        \psi(s)=\frac{\mathrm{L}}{\sqrt{-\Lambda_m}\,\kappa^2}\,\tanh \left(\sqrt{-\Lambda_m}\,s\right)
        \hfill\\  [2ex]
        t(s)=\frac{\mathrm{L}}{(-\Lambda_m)^{3/2}\,\beta\,\kappa^2}\,\tanh \left(\sqrt{-\Lambda_m}\,s\right)
    \end{array}$\\[6ex]
    \hline 
\end{tabular}
\caption{Time-like Geodesic equations for massless BTZ space-time}
\label{tab:3}
\end{table}

From the above geodesic analysis on massless BTZ space-time, it is evident the modified gravity theories (Ricci-Inverse and $f(\mathcal{R})$ gravity) influences the geodesic paths and shifted the result compared to general relativity case. 

\subsection{ Geodesic Equation in $AdS_3$-type BTZ space-time in RI-gravity }\label{subsec:2}

In this part, we discuss a vacuum space-time of BTZ metric obtained in Ricci-Inverse (RI) gravity given in Eq. (\ref{special-metric}) under the scenario of $M=-1$ and zero angular momentum, $J=0$.

Therefore, from space-time (\ref{special-metric}), we find a $Ads_3$-type space-time in modified gravity given by
\begin{equation}
    ds^2=-(1-\Lambda_m\,r^2)\,dt^2+\frac{dr^2}{(1-\Lambda_m\,r^2)}+r^2\,d\psi^2.\label{special-metric4}
\end{equation}

\begin{center}
\begin{figure}[ht!]
\begin{centering}
\subfloat[$\Lambda=-0.1$]{\centering{}\includegraphics[scale=0.6]{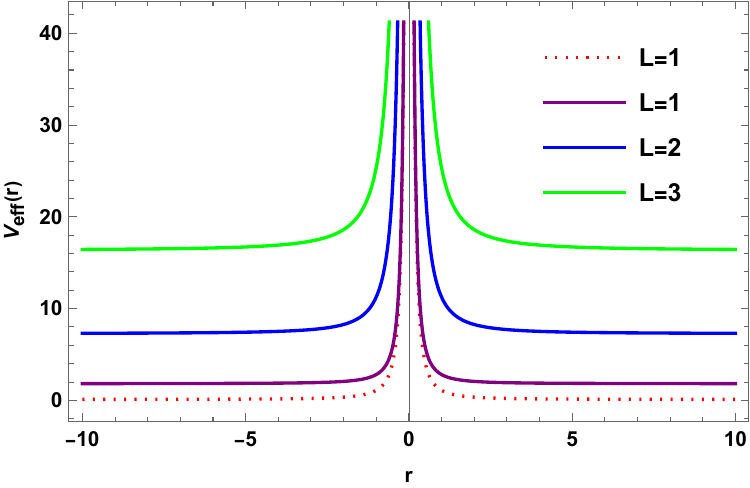}}\quad\quad
\subfloat[$\Lambda=-0.1$,$\mathrm{L}=1$]{\centering{}\includegraphics[scale=0.6]{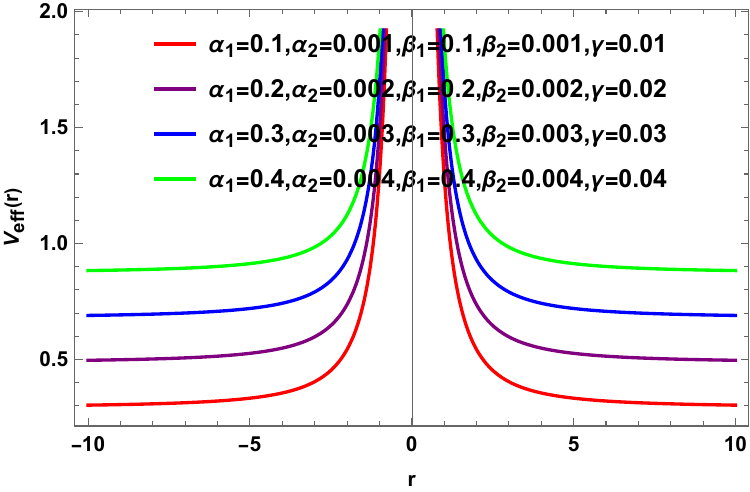}}\quad
\centering{}\caption{The effective potential for null geodesics on $AdS_3$-type BTZ space-time in $\mathcal{RI}$-gravity.}\label{fig:9}
\hfill\\
\subfloat[$\Lambda=-0.1$]{\centering{}\includegraphics[scale=0.6]{time-like-geodesics-massless-figure-1-a.pdf}}\quad\quad
\subfloat[$\Lambda=-0.1$,$\mathrm{L}=1$]{\centering{}\includegraphics[scale=0.6]{time-like-geodesics-massless-figure-1-b.pdf}}\quad
\centering{}\caption{The effective potential for time-like geodesics on $AdS_3$-type BTZ space-time in $\mathcal{RI}$-gravity.}\label{fig:10}
\par\end{centering}
\end{figure}
\par\end{center}

As done in the previous analysis, the geodesics equations for the space-time (\ref{special-metric4}) are given by
\begin{eqnarray}
    \dot{t}=\frac{\mathrm{E}}{h(r)},\quad h(r)=(1-\Lambda_m\,r^2),\quad\quad
    \dot{\psi}=\frac{\mathrm{L}}{r^2},\quad\quad
    \dot{r}^2+\left(-\epsilon+\frac{\mathrm{L}^2}{r^2}\right)\,h(r)=\mathrm{E}^2.\label{B1}
\end{eqnarray}
In that scenario, the effective potential of the system either null $(\epsilon=0)$ or time-like $(\epsilon=-1)$ geodesics is given by
\begin{equation}
    V_{eff} (r)=\left(-\epsilon+\frac{\mathrm{L}^2}{r^2}\right)\,(1-\Lambda_m\,r^2).\label{B2}
\end{equation}
We see that the effective potential of the system in this scenario $M=-1$ and $J=0$ depends on the coupling constant $\alpha_i,\beta_i$ and $\gamma$.

\begin{center}
\begin{figure}[ht!]
\begin{centering}
\subfloat[$\Lambda=-0.1$]{\centering{}\includegraphics[scale=0.6]{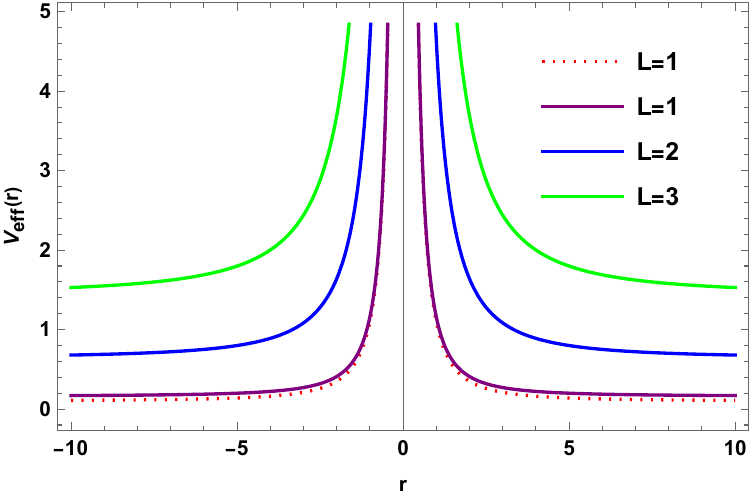}}\quad\quad
\subfloat[$\Lambda=-0.1$,$\mathrm{L}=1$]{\centering{}\includegraphics[scale=0.6]{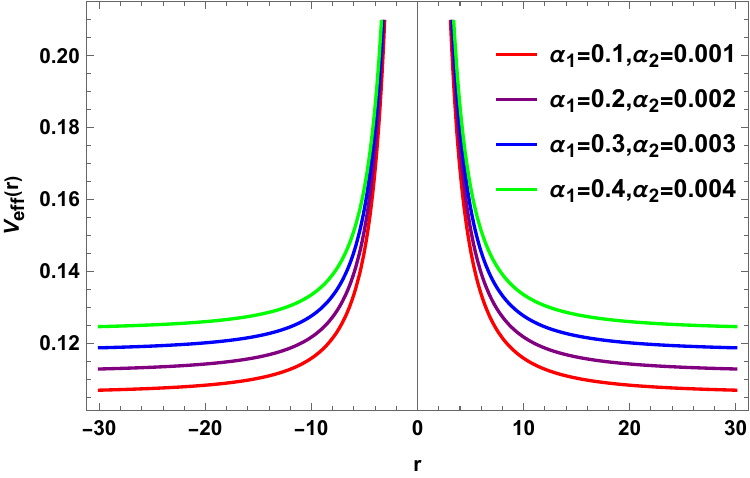}}\quad
\centering{}\caption{The effective potential for null geodesics on $AdS_3$-type BTZ space-time in $f(\mathcal{R})$ gravity.}\label{fig:11}
\hfill\\
\subfloat[$\Lambda=-0.1$]{\centering{}\includegraphics[scale=0.6]{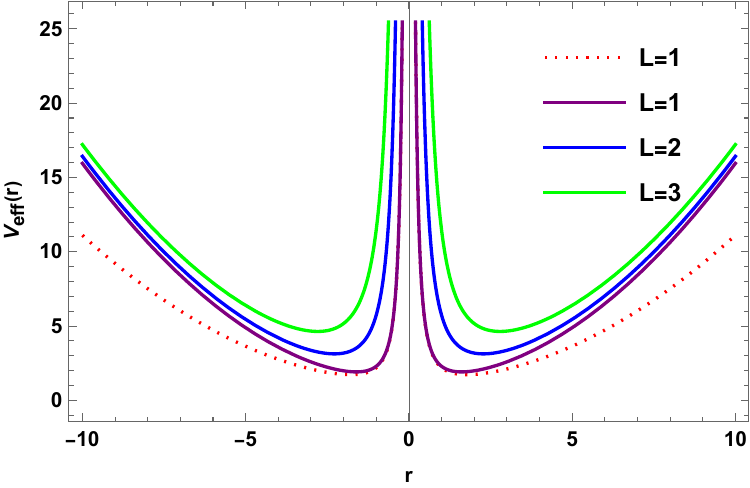}}\quad\quad
\subfloat[$\Lambda=-0.1$,$\mathrm{L}=1$]{\centering{}\includegraphics[scale=0.6]{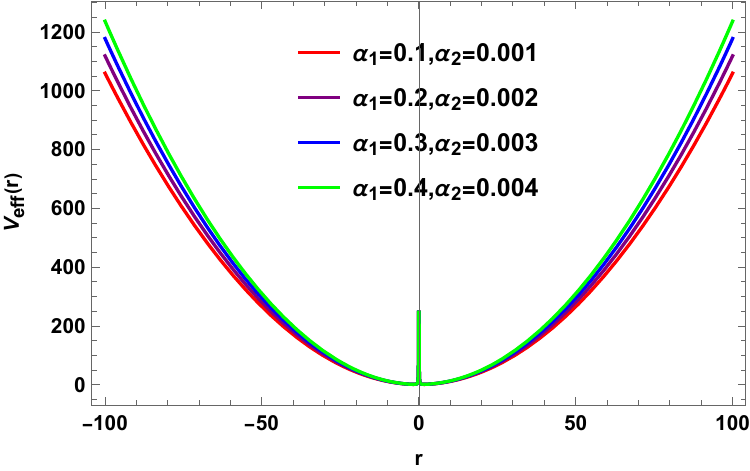}}\quad
\centering{}\caption{The effective potential for null geodesics on $AdS_3$-type BTZ space-time in $f(\mathcal{R})$ gravity.}\label{fig:12}
\par\end{centering}
\end{figure}
\par\end{center}

We have generated Figure \ref{fig:9}, which illustrates the behavior of the effective potential for null geodesics in $AdS_3$-type BTZ space-time within the framework of Ricci-Inverse gravity and General Relativity (GR). In panel (a) of Figure \ref{fig:9}, the dotted red line representing $\mathrm{L}=1$ corresponds to the GR case, where the parameters are set to $\alpha_i=0$, $\beta_i=0$, and $\gamma=0$. In contrast, the purple line for $\mathrm{L}=1$ represents the Ricci-Inverse gravity scenario, with the coupling constants set as $\alpha_1=0.1$, $\alpha_2=0.001$, $\beta_1=0.1$, $\beta_2=0.001$, and $\gamma=0.01$. The other lines in this panel depict different values of 
$\mathrm{L}$ while maintaining the same coupling constants. In panel (b) of Figure \ref{fig:9}, the lines correspond to various values of the coupling constants while keeping $\mathrm{L}=1$ fixed. This setup allows for a comparative analysis of the effective potential for null geodesics in $AdS_3$-type BTZ space-time within the GR and Ricci-Inverse gravity theories.

Similarly, we have generated Figure \ref{fig:10} illustrating the behavior of the effective potential for time-like geodesics in $AdS_3$-type BTZ space-time within the Ricci-Inverse gravity and General Relativity (GR). Here also, the explanation is analogous to the previous Figure \ref{fig:9}. 

In Figure \ref{fig:11} and (\ref{fig:12}), we have depicted the effective potential for null and time-like geodesics in $AdS_3$-type BTZ space-time within the framework of $f(\mathcal{R})$ gravity and show how the coupling constants influences on them.

\section{Conclusions}\label{sec:5}

In the current study, we investigated a stationary BTZ space-time within the framework of Ricci-Inverse gravity, a novel approach to modified gravity theories, where the inverse of the Ricci tensor plays a central role in the dynamics. We considered a general class of this theory characterized by the function $ f(\mathcal{R}, \mathcal{A}, A^{\mu\nu}A_{\mu\nu})$ and demonstrated that the stationary BTZ space-time emerges as a valid solution in this theory, as discussed in section \ref{sec:2}. Notably, this solution shows modification to the cosmological constant, influenced by the coupling constants within this theory, thereby enhancing our understanding of solution in modified gravity frameworks. Additionally, we explored another modified gravity theory known as $f(\mathcal{R})$-gravity in section \ref{sec:3}. For that, we have chosen the function $f$ to the form $f(\mathcal{R})=(\mathcal{R}+\alpha_1\,\mathcal{R}^2+\alpha^2\,\mathcal{R}^2)$ into the Einstein-Hilbert action. Considering this BH space-time (\ref{1}) as background model, we solved the modified field equations in a vacuum, and obtained the result in $f(\mathcal{R})$-gravity. Our analysis showed that the coupling constants associated with $\mathcal{R}^2$ and $\mathcal{R}^3$ induces modification to the cosmological constant while preserving its negative nature. 

One of the central contributions of this paper is the complete integration of the geodesic equations for the BTZ space-time within the contexts of Ricci-Inverse gravity and general relativity which was discussed in section \ref{sec:4}. We expressed these solutions in terms of elementary functions, providing explicit forms for the trajectories of both massive and massless test particles. We classified the geodesics-both null and time-like-under various scenarios: in the general case where $M \neq 0$ and $J \neq 0$, and in the degenerate cases, such as $M=-1$ and $J=0$ in subsection \ref{subsec:1}, as well as the scenario where both mass and angular momentum vanish, $M=0=J$, discussed in subsection \ref{subsec:2}. We have shown in each section that the results get influenced by the modified gravity theory compared to general relativity case.

For future work, it would be valuable to extend the analysis of BTZ black holes within the Ricci-Inverse and $f(\mathcal{R})$ gravity framework by exploring their thermodynamic properties and stability under different perturbations. This could include studying quantities such as heat capacity and entropy to gain deeper insights into potential phase transitions within the context of this modified gravity theory. 

\section*{Acknowledgments}
F.A. acknowledges the Inter University Centre for Astronomy and Astrophysics (IUCAA), Pune, India for granting visiting associate-ship. 

\section*{Data Availability Statement}

This manuscript has no associated data.

\section*{Conflict of Interest}

Author(s) declare no such conflict of interests.

\section*{Code/Software}

This manuscript has no associated code/software.

\end{document}